\documentclass[prl,twocolumn,superscriptaddress,preprintnumbers,showpacs,byrevtex,floatfix]{revtex4}
\usepackage{graphicx}
\usepackage{amssymb}
\usepackage{array}
\usepackage{verbatim}
\usepackage[ansinew]{inputenc}
\usepackage[english]{babel}
\usepackage{bm}
\usepackage{natbib}
\usepackage{hyperref}
\usepackage{amsmath}
\usepackage{multirow}

\begin{document}
\DeclareGraphicsExtensions{.pdf,.png,.jpg,.eps,.tiff}
\title{Versatile ytterbium ion trap experiment for operation of scalable ion trap chips with motional heating and transition frequency measurements}
\author{James J. McLoughlin}
\author{Altaf H. Nizamani}
\author{James D. Siverns}
\author{Robin C. Sterling}
\author{Marcus D. Hughes}
\author{Bjoern Lekitsch}
\author{Bj\"orn Stein}
\author{Seb Weidt}
\author{Winfried K. Hensinger \footnote{W. K. Hensinger@sussex.ac.uk\\ URL: http://www.sussex.ac.uk/physics/iqt}}
\affiliation{Department of Physics and Astronomy, University of Sussex, Brighton, BN1 9QH, UK}

\begin{abstract}


We present the design and operation of an ytterbium ion trap experiment with a setup offering versatile optical access and 90 electrical inter-connects that can host advanced surface and multi-layer ion trap chips mounted on chip carriers. We operate a macroscopic ion trap compatible with this chip carrier design and characterise its performance, demonstrating secular frequencies $>$1 MHz, and trap and cool nearly all of the stable isotopes, including $^{171}$Yb$^{+}$ ions, as well as ion crystals. For this particular trap we measure the motional heating rate, $\langle \dot{n} \rangle$, and observe a $\langle \dot{n} \rangle \propto 1/\omega^{2}$ behaviour for different secular frequencies, $\omega$. We also determine a spectral noise density $S_{\text{E}}(1$ MHz) = 3.6(9)$\times10^{-11}$ V$^{2}$m$^{-2}$Hz$^{-1}$ at an ion electrode spacing of 310(10) $\mu m$. We describe the experimental setup for trapping and cooling Yb$^{+}$ ions and provide frequency measurements of the $^{2}S_{1/2}$$\leftrightarrow$$^{2}P_{1/2}$ and $^{2}D_{3/2}$$\leftrightarrow$$^{3}D[3/2]_{1/2}$ transitions for the stable $^{170}$Yb$^{+}$, $^{171}$Yb$^{+}$, $^{172}$Yb$^{+}$, $^{174}$Yb$^{+}$ and $^{176}$Yb$^{+}$ isotopes which are more precise than previously published work.

\end{abstract}
\maketitle

\setcounter{secnumdepth}{1}
\section{Introduction}
Ions confined in radio-frequency (rf) traps are regarded as one of the most promising approaches towards quantum information processing \cite{zoller, wineland, haffner}, as well as for quantum simulators \cite{pons, friedenaueri, Johanning, clarki, kim} and frequency standards \cite{udem, Webster, Tamm, chwalla}. Their suitability has been well demonstrated with state preparation and detection \cite{schaetzi, actoni, Wunderlich, myersoni, burrell2}; qubit entanglement, gates and error correction \cite{sacketti, leibfriedai, schmidti, chiaverinii, brickman, home, Ospelkaus, monz, wang, timoney}; and transport of ions within ion trap arrays \cite{rowe, hensinger, schulz1, reichle, Pearson, Hucul, Huber, Blakestad, Aminiz}. Development of scalable traps that can encompass all of these operations is now the next stage towards more practical systems incorporating microfabricated chip traps \cite{stick, chiaverini2, Brownnutt, Brown, Britton, steane, Seidelin, Labaziewicz, Britton2, Leibrandt, allcock, Aminiz}, but realising experimental setups can be challenging. We have developed an experiment honed towards the development of trap architectures which includes a vacuum system that accommodates both surface and multi-layer traps and provides up to 90 electrical connections enabling the design and testing of a wide range of complex trap geometries to be mounted on commercial compatible chip carriers. We discuss the experimental setup, particularly describing the vacuum system, laser locking and imaging devices used. We describe a macroscopic ion trap compatible with chip carrier mounting and characterise its operation. Noise such as electric field fluctuations (from thermal electric Johnson-Nyquist noise) and fluctuating patch potentials on electrodes induces motional heating of trapped ions \cite{Turchette, Deslauriers2, Labaziewicz, Labaziewicz2}. We measure the motional heating rate within our ion trap using a technique where the laser cooling beam is turned off for a certain time and ion fluorescence is monitored when the laser is turned on again \cite{NIST1, NIST2}. There are currently several ions being investigated for quantum technology including Ba$^{+}$ \cite{Dietrich}, Be$^{+}$ \cite{wineland2}, Ca$^{+}$ \cite{nagerl, Lucas, hughes, thompson, schulz}, Cd$^{+}$ \cite{blinov}, Mg$^{+}$ \cite{barrett}, Sr$^{+}$ \cite{Pearson, Letchumanan}, and Yb$^{+}$ \cite{bell, balzer, kielpinski, olmschenk}. Yb$^{+}$ is an attractive choice for quantum information processing and for frequency standards \cite{roberts, gill2, schneider, kazumoto}. We present frequency measurements more precise than previously published results \cite{streed} for the $^{2}S_{1/2}$$\leftrightarrow$$^{2}P_{1/2}$ transition and the $^{2}D_{3/2}$$\leftrightarrow$$^{3}D[3/2]_{1/2}$ for the Yb$^{+}$ isotopes: $^{170}$Yb$^{+}$, $^{171}$Yb$^{+}$, $^{172}$Yb$^{+}$, $^{174}$Yb$^{+}$, $^{176}$Yb$^{+}$.

\section{Experimental Setup}
\begin{figure}[!h]
\centering
\includegraphics[scale=0.18] {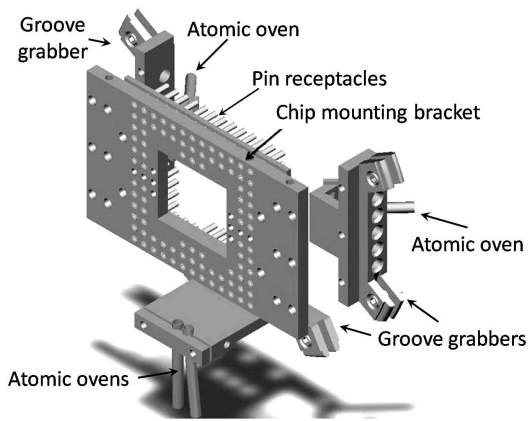}
\caption{View of the mounting bracket showing the 90 pin receptacles, and positions of the four atomic ovens. Two ovens are behind the pins and angled to provide ytterbium atoms to multi-layer ion traps featuring a through-hole. The other two ovens provide atoms for traps where access is parallel to the ion trap surface.}
\label{chipcarrier}
\end{figure}

Our vacuum system has some similar features as the one used in ref. \cite{stick} and includes a custom mounting bracket with 90 electrical connections, four atomic ovens and versatile optical access to cater for traps that require laser access via a through hole and traps that require laser beams parallel to the surface of the trap. The mounting bracket, shown in Fig. \ref{chipcarrier}, consists of two ultra-high vacuum (UHV) compatible polyether ether ketone (PEEK) plates sandwiching 90 gold-plated receptacles (Mill-Max; 0672-1-15-15-30-27-10-0) which are arranged to be compatible with 101-pin CPGA chip carriers (Global Chip Material; part no. PGA10047002). This setup enables complex trap structures that are mounted onto chip carriers to be easily inserted into the mounting bracket and changed with minimal turn around time. Kapton insulated copper wires connect 88 receptacles to two 50-pin feedthroughs providing control of the ion trap static voltage electrodes. Two receptacles are used for RF and grounding and are connected to a power feedthrough (Kurt J. Lesker; EFT0521052) with a pair of short thick bare copper wires insulated with ceramic beads.

\begin{figure}[!h]
\centering
\includegraphics[scale=0.18] {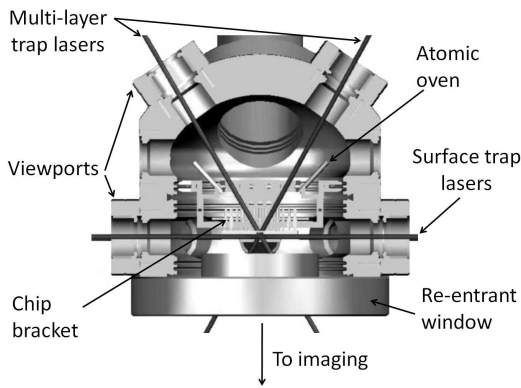}
\caption{Top cut-view of the vacuum chamber (consisting of an octagon and hemisphere flange) showing the mounting bracket, atomic ovens and laser access for symmetric and surface traps.}
\label{chamber}
\end{figure}

Groove grabbers attach the mounting bracket to a vacuum chamber, comprising of an octagon and hemisphere (Kimball Physics, part nos. MCF450-SO20008-C and MCF450-MH10204/8-A respectively), as shown in Fig. \ref{chamber}. Four atomic ovens (ohmically heated stainless steel tubes) are arranged in the chamber to provide the surface and multi-layer traps with either natural ytterbium or an enriched source of $^{171}$Yb atoms. Eight 1.33" conflat (CF) mounted anti-reflection (AR) coated fused quartz silica viewports allow laser access to the centre of both trap types, and a specially designed 4.5" CF mounted AR coated re-entrant window is mounted in front of the chamber to allow imaging of the ion. The distance between the ion and window surface is 4 mm. A UHV environment of $10^{-12}$ torr is accomplished using a $20$ ls$^{-1}$ ion pump (Varian; part no. 9191145) and titanium sublimation pump (Varian; part no. 9160050).

The RF drive frequency is generated using a HP8640B signal generator, amplified (NP Technologies no. NP-541) and passed through a resonator \cite{Macalpiner, siverns} which is then attached to the power feedthrough. The resonator design specifications are detailed in table \ref{res} and when loaded with the trap has a quality factor $Q = 200(20)$, and geometric factor $\kappa = 24(8)$, defined by $V = \kappa\sqrt{PQ}$ where $V$ is the RF voltage applied to the electrodes, $P$ the power applied to the electrodes and $Q$ the quality factor of the resonator \cite{siverns}. Impedance matching is achieved by measuring the percentage of reflected power to applied power using a directional power-meter (Rhode and Schwarz; part no. NAUS 3), and the coupling is maintained at $\geq95\%$.

\begin{table}[ht]
	\centering
        \caption{Specifications of the resonator}
		\begin{tabular}{cp{36mm}c}
            \hline
            \hline
            \multirow{10}{*}{\includegraphics[scale=0.12] {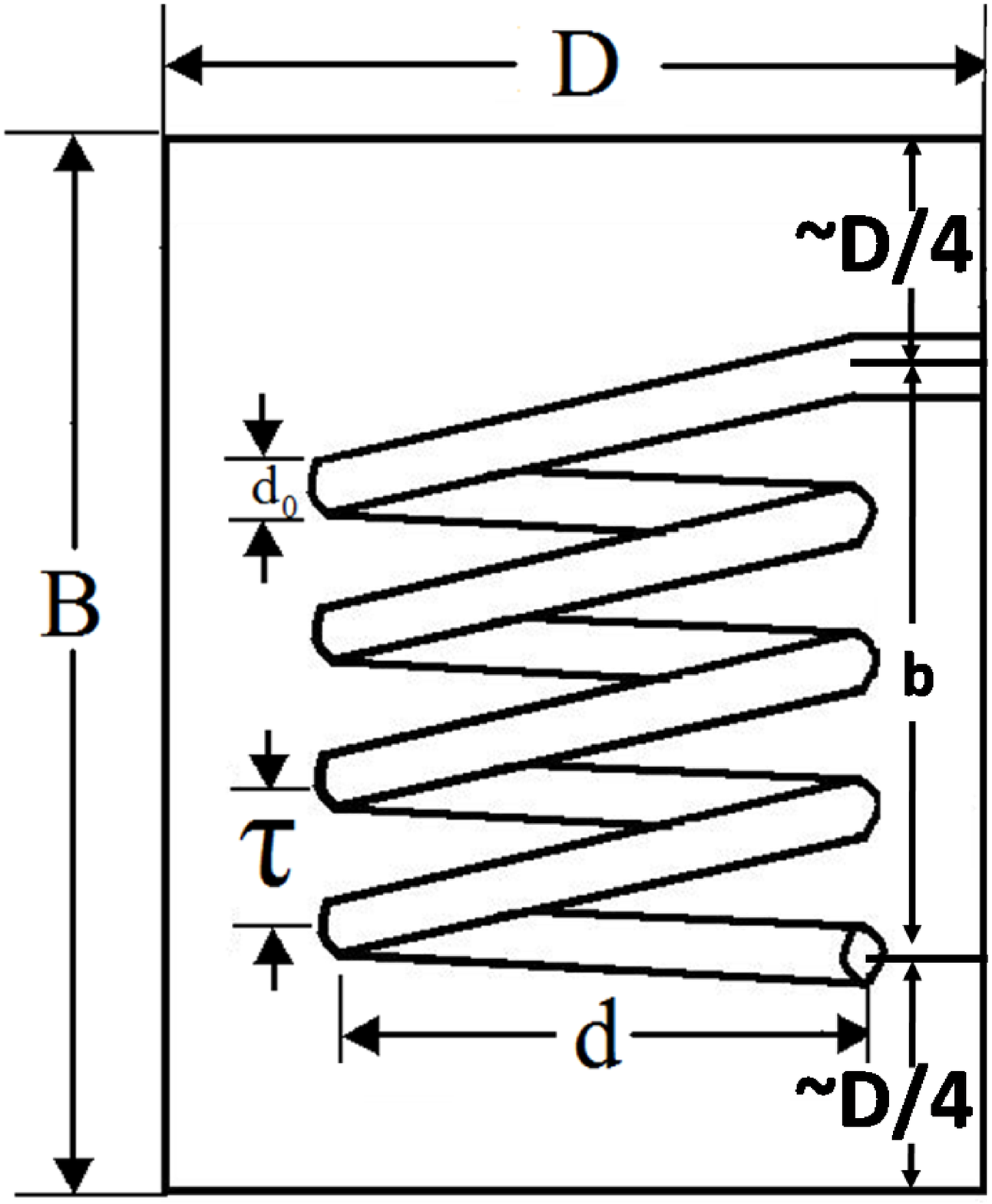}} & Shield Diameter, D & $76(1)$ mm\\
            & Shield length, B & $103(1)$ mm\\
            & Coil Diameter, $d$ & $52(3)$ mm\\
            & Coil Length, $b$ & $63(5)$ mm\\
            & Coil diameter, $d_0$ & $3.14(3)$ mm\\
            & Winding pitch, $\tau$ & $6(2)$ mm\\
            & Number of turns, $N$ & $9.50(25)$ \\
            & Resonant frequency with trap load, $f_0$ & $21.5(1)$ MHz\\
            & Q with trap load& $200(20)$\\ \hline \hline
        \end{tabular}
	\label{res}
\end{table}

\begin{figure}[!h]
\centering
\includegraphics[scale=0.25]{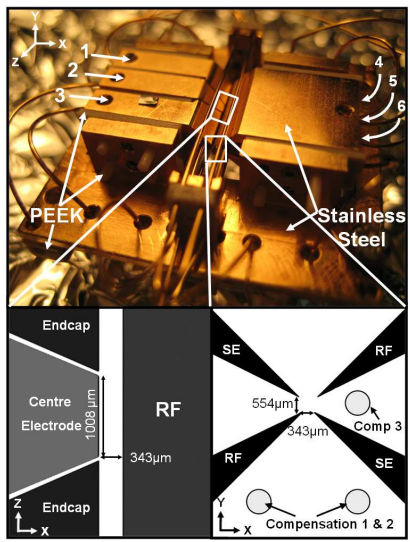}
\caption{(Color online) Top: Ion trap used in the experiments. Bottom right: Cross-sectional view of the radial plane of the trap, showing the spacings between the radio frequency (RF) electrodes and the static voltage electrodes (SE). Bottom left: View of the trap showing the length of the centre electrode and trapping region.}
\label{trap}
\end{figure}

The macroscopic linear RF-Paul trap used is shown in Fig. \ref{trap}. To be compatible with the vacuum system it is limited to a footprint of 3 cm x 3 cm and a height of 1 cm, and can be plugged into the mounting bracket. A PEEK base is used to avoid electrical shorting with the mounting bracket connections. Exposure to laser radiation has been observed to increase outgassing and induce discolouring of PEEK. This is suspected to cause charging of the dielectric, so a stainless-steel mount is used to shield the ion from any exposed dielectrics. Blade shaped electrodes, electroplated with 5 $\mu m$ of gold, are suspended from the stainless steel mount. The RF electrodes span the entire axial length of the trap, while the static voltage electrodes are segmented to provide end cap potentials and rotation of the principal axes. This arrangement provides for a reduced residual RF ponderomotive potential along the $z$ axis \cite{Madsen} to $<2 \%$ of the radial frequency, and allows trapping of long ion chains without appreciable RF micromotion in the axial direction. The electrodes are separated by $343(14)$ $\mu m$ on the $x$ axis and $554(14)$ $\mu m$ on the $y$ axis, producing a trap aspect ratio of 1.6, and the ion-electrode separation is $310(10)$ $\mu m$. The length of the centre electrode is $1008(14)$ $\mu m$. Three parallel wires, compensation electrodes $1-3$, run the length of the trap to provide micromotion compensation. Holes in the mount coincide with respective receptacles in the mounting bracket and gold plated wires connect the electrodes to the pin receptacles.

\begin{figure}[!h]
\centering
\includegraphics[scale=0.35] {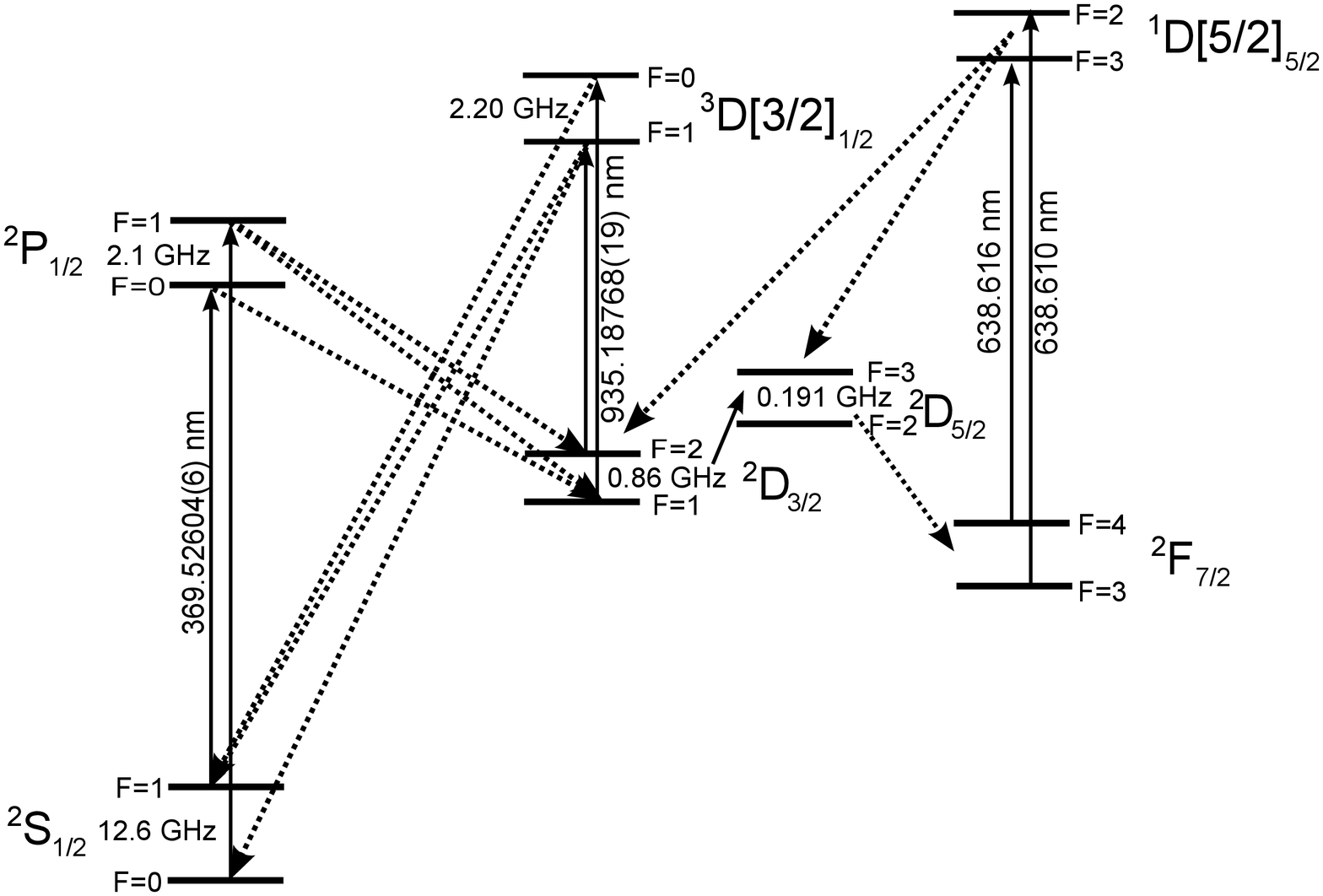}
\caption{Energy level scheme for $^{171}$Yb$^{+}$, where the ground state hyperfine levels $^{2}S_{1/2}|$F=0,$m_{F}$=0$\rangle$ and $^{2}S_{1/2}|$F=1,$m_{F}$=0$\rangle$ represent the qubit states $|0\rangle$ and $|$1$\rangle$ respectively. The $^{2}S_{1/2} \leftrightarrow ^{2}P_{1/2}$, used for Doppler cooling and fluorescence detection, is driven by a 369.5 nm beam with 14.7 GHz frequency sidebands. Decay from the $^{2}P_{1/2}$ state to the metastable $^{2}D_{3/2}$ state is returned to the ground state by exciting the $^{2}D_{3/2}$ $\leftrightarrow$ $^{3}D[3/2]_{1/2}$ transition using a 935.2 nm beam with 3.08 GHz frequency sidebands. The $^{2}F_{7/2}$ state is populated a few times per hour and the ion is returned to the ground state using 638.6 nm laser switched between the two $^{2}F_{7/2}$ $\leftrightarrow$ $^{1}D[5/2]_{5/2}$ transitions. Wavelength values (vacuum) are shown.}
\label{171yb}
\end{figure}

To create Yb$^{+}$ ions neutral atoms are directed towards the centre of a trapping region and ionised using two-colour photoionisation \cite{balzer}. Here a 398.9 nm photon excites the neutral $^{1}S_{0}$$\leftrightarrow$$^{1}P_{1}$ transition, with a following 369.5 nm photon ionising the atom. Fig. \ref{171yb} shows the energy level diagram for $^{171}$Yb$^{+}$, with the hyperfine doublets resulting from the ions spin-1/2 nucleus. The ground state levels, $^{2}S_{1/2}|$F = 0,$m_{\text{F}}$ = 0$\rangle$ and $^{2}S_{1/2}|$F = 1,$m_{\text{F}}$ = 0$\rangle$, represent the qubit states $|0\rangle$ and $|$1$\rangle$ respectively. The stable isotopes $^{170}$Yb$^{+}$, $^{172}$Yb$^{+}$, $^{174}$Yb$^{+}$ and $^{176}$Yb$^{+}$ have zero nuclear spin, resulting in similar energy level structures but without a hyperfine structure.

$^{171}$Yb$^{+}$ ions are Doppler cooled on the 369.5 nm $^{2}S_{1/2}$$\leftrightarrow$$^{2}P_{1/2}$ electric dipole transition (natural linewidth $\Gamma/2\pi= 19.6$ MHz \cite{Olmschenk3}). The 369.5 nm beam, with 14.7 GHz frequency sidebands used to repump hyperfine states, is generated by frequency doubling a 7.35 GHz phase modulated (New Focus electro-optic modulator: 4851) 739.05 nm beam in a resonant doubling cavity (Toptica Photonics TA-SHG with free spectral range of 1.05 GHz).

The $^{2}P_{1/2}$ state decays (branching ratio $\approx0.005$ \cite{olmschenk}) to the metastable $^{2}D_{3/2}$ state (lifetime $\tau$ = 52 ms \cite{Engelke}). A 935.2 nm laser, current modulated at 3.08 GHz using a bias-t (Mini Circuits; part no. ZFBT-4R2G+), excites transitions to the $^{3}D[3/2]_{1/2}$ state allowing rapid population of the ground state. Inelastic collisions with background gas ($\approx$ 30 mins) results in population of the $^{2}F_{7/2}$ state (natural lifetime of $\approx6$ years \cite{kazumoto}). A laser scanned between 638.610 nm and 638.616 nm \cite{roberts} (by adjusting the laser current and grating angle simultaneously at a rate of 10 Hz) excites transitions to the $^{1}D[5/2]_{5/2}$ state which decays back to the $^{2}D_{3/2}$ state. The 398.9 nm, 638.6 nm, and 935.2 nm lasers systems are constructed in-house in the Littrow configuration.

\begin{figure}[!h]
\centering
\includegraphics[scale=0.2]{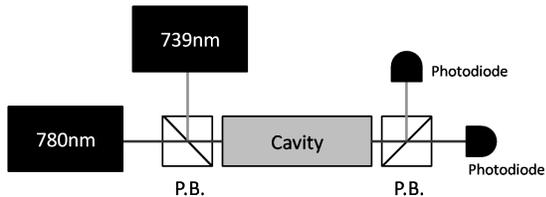}
\caption{Pictorial representation of the optical setup for a scanning confocal cavity. The lasers are combined using a polarising beam splitter and aligned into the cavity. The exiting light is then separated with another polarising beam splitter onto two photodiodes. The photodiode signals are sampled using a National Instruments analog input sampling card (National Instruments PCI-6143).}
\label{cavity}
\end{figure}

\begin{figure}[!h]
\centering
\includegraphics[scale=0.3]{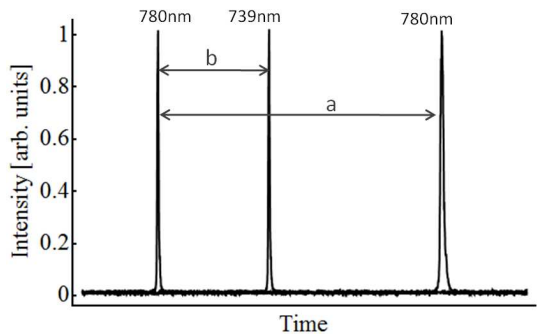}
\caption{The combined signal from both photodiodes. The free spectral range of the 780 nm reference laser, a, is measured as well as the separation between the first 780 nm peak and the first 739.05 nm peak, b. These times are compared and the resulting ratio is maintained to stabilise the 739.05 nm laser.}
\label{peaks}
\end{figure}

Efficient cooling and repumping requires that the lasers are frequency stabilised to within the linewidth of the respective transitions. The 739.05 nm and 935.2 nm lasers are locked to a stabilised 780 nm laser using a transfer cavity locking scheme \cite{comp-based}. The 780 nm laser, locked to the $^{87}$Rb D2 transition to $<$1 MHz accuracy, is directed into two scanning confocal Fabry-Perot cavities each with a finesse of 134(5) and free spectral ranges of 1 GHz and 750 MHz for the 780/739 nm and 780/935 nm cavities respectively. The 739.05 nm and 935.2 nm lasers are then each directed into one of these cavities. Fig. \ref{cavity} illustrates the scheme for the 739.05 nm laser. The resonance peaks from the cavity are read into a computer using an analog input card (National Instruments PCI-6143). The cavities are scanned over 4 GHz at a rate of 100 Hz, and combined with the sample rate of the input card, produces a resolution of 3 MHz so peaks with FWHMs of 40 MHz consist of $\approx$ 15 samples. Using LabVIEW Realtime the free spectral range of the 780 nm peaks, $a$, is compared to the separation of 780 nm and 739.05 nm (935.2 nm) resonance peaks, $b$, as shown in Fig. \ref{peaks}. A change in the ratio generates an error signal which is sent, via a NI PCI-6722 analog output card, to the piezo controlling the grating angle on the 739.05 nm (935.2 nm) laser. Feeding back to the cavity piezos to hold the position of the first peak constant with respect to the scan voltage stabilises the cavity length and removes the requirement for thermal isolation. This technique allows wavelengths to be tuned over a range of $\approx$ 500 MHz via computer control.

To drift compensate the 398.91 nm and 638.6 nm lasers the wavelengths are measured (High Finesse WS7, accurate to 60 MHz) and read into a LabVIEW program. The program generates error signals that are sent to the lasers, via the output card (National Instruments: PCI-6722), maintaining the wavelengths.

\begin{figure}[!h]
\centering
\includegraphics[scale=0.23]{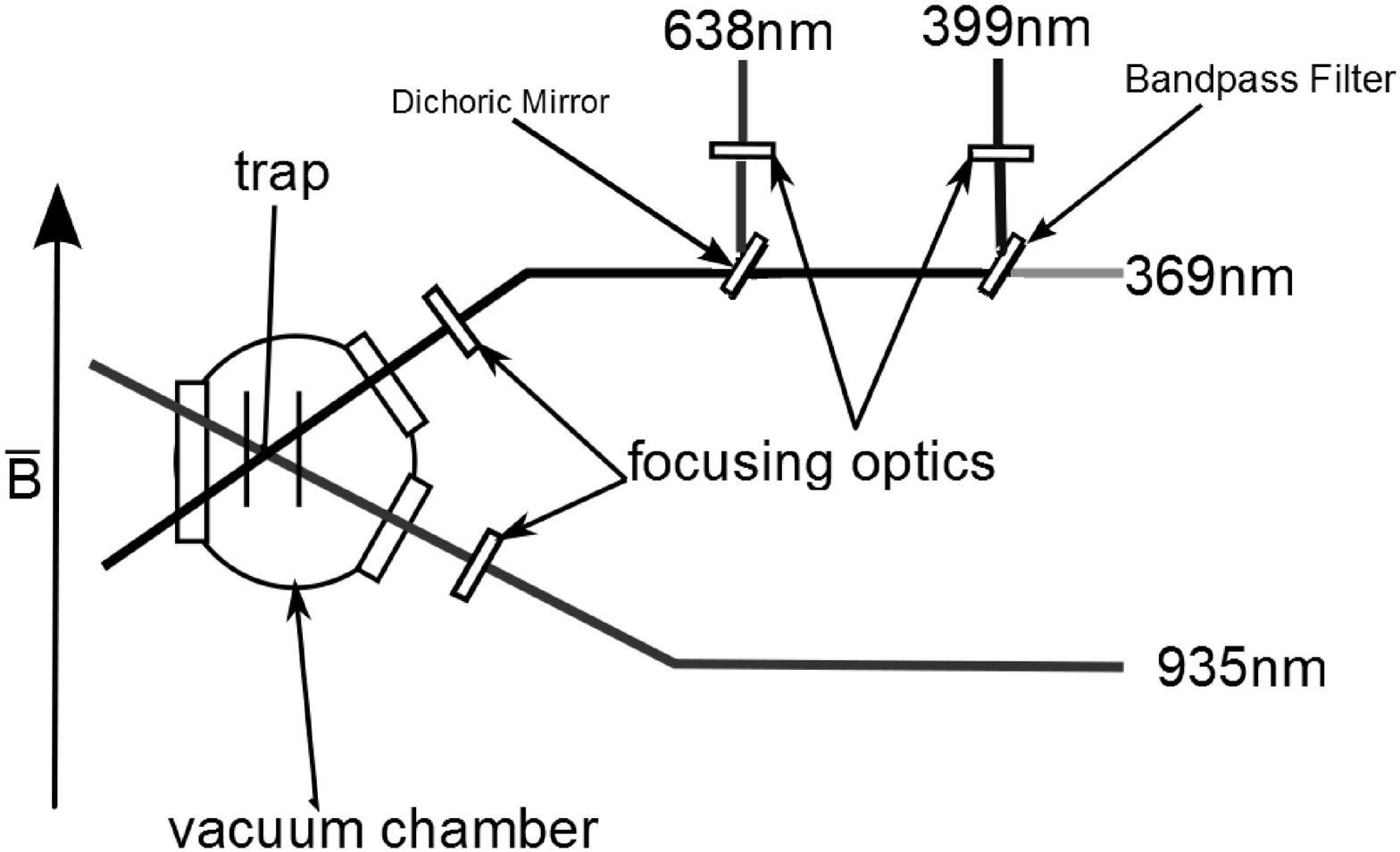}
\caption{Diagram showing basic laser set up to provide a good overlap of all four laser beams through the trapping centre. A bandpass filter (Semrock: FF01-370/36-25) is used to combine the 369.5 nm and 398.9 nm beams, which are then combined with the 638.6 nm beam using a dichroic mirror. Also shown is the 0.5 mT applied magnetic field, used to define the quantisation axis. For $^{171}$Yb$^{+}$ the lasers are polarised at oblique angles to the quantisation axis to avoid potential dark states.}
\label{laserset}
\end{figure}

The stabilised lasers are aligned into the trap as illustrated in Fig. \ref{laserset}. The 398.91 nm and 369.5 nm beams, with powers $\approx$ 500 $\mu$W and $\approx$ 1 mW respectively, are focused to beam waists of 30(5) $\mu m$. The higher powers of the 638.6 nm and 935.2 nm beams, $\approx$ 7 mW each, allow for wider foci of 60(5) $\mu m$ while maintaining sufficiently high beam intensities. The beam polarisations are orientated at 45$^{o}$ with respect to a 0.5 mT externally applied magnetic field (defining the quantisation axis) to couple all Zeeman hyperfine levels to the cooling cycle.

Ions are detected by measuring fluorescence using either an electron multiplied CCD array (EMCCD), Andor: iXon885 or a photo multiplier tube (PMT), Hamamatsu: H8259-01. The device chosen depends upon the the experiment being performed, desired detection accuracy and whether spatial resolution is required.

\section{Ion trap performance}
With this experimental setup we can trap single ions and ion crystals of most of the stable Yb$^{+}$ isotopes, including $^{171}$Yb$^{+}$. To trap ions the static and rf voltages shown in table \ref{voltages} are applied to the electrodes, corresponding to the numbered electrodes in Fig. \ref{trap}, along with an RF drive signal with a frequency of $\Omega/2\pi = 21.48$ MHz.

\begin{table}[!h]
\begin{center}
\caption{Voltages applied to the trap electrodes, numbered in Fig. \ref{trap}.}
\begin{tabular}{cc}
    \hline \hline
    Electrode & Voltage [V]\\ \hline
    1 & 148.88 (1) \\
    2 & 7.36 (1) \\
    3 & 25.03 (1) \\
    4 & 0.00 (1) \\
    5 & 0.00 (1) \\
    6 & 167.76 (1) \\
    compensation 1 & 169.22 (1) \\
    compensation 2 & 169.22 (1) \\
    compensation 3 & -2.70 (1) \\
    RF & 680 (10) \\ \hline
    \hline
\end{tabular}

\label{voltages}
\end{center}
\end{table}

\begin{figure}[!h]
\centering
\includegraphics[scale=0.4] {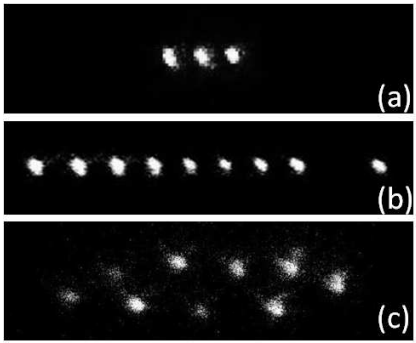}
\caption{Crystal of (a) three Yb$^{+}$ ions, (b) multi ion chain of mixed species, and (c) a zig-zag structure of nine Yb$^{+}$ ions}
\label{chain}
\end{figure}

The ion secular frequencies, ($\omega_{x}$, $\omega_{y}$, $\omega_{z}$)/2$\pi$, are measured to  be (2.069, 2.110, 1.030) $\pm$0.001 MHz respectively. They are found by applying an AC voltage to one of the endcap electrodes and scanning its frequency. When equal to one of the secular frequencies, the ion is resonantly heated leading to a visible de-crystallisation. Using these measured secular frequencies, along with electric field simulations \cite{Hucul} the trap depth is determined to be 4.9(2) eV. Ion lifetimes of many hours, without optical cooling, have been observed and ion crystals, as shown in Fig. \ref{chain}, can be produced. Fig. \ref{chain}(a) shows a three ion crystal, $(b)$ a multi-isotope crystal highlighting the potential for many isotope experiments such as sympathetic cooling, and $(c)$ a zig-zag ion crystal.

\section{Frequency measurements}
Using trapped Yb$^{+}$ ions the exact cooling and repumping wavelengths are determined. The wavelengths required to excite the neutral $^{1}S_{0}$$\leftrightarrow$$^{1}P_{1}$ transition, creating Yb$^{+}$ ions, are determined using the fluorescence spot method described in \cite{Nizamani}. The Doppler cooling and repumping wavelengths are determined by scanning the wavelength of the respective laser while monitoring the ion fluorescence, and measuring the wavelength on a wavemeter (High Finesse: WS7). The intensity of the 369.5 nm and 935.2 nm beams are reduced to 0.4 Wcm$^{-2}$ was 0.02 Wcm$^{-2}$ respectively to reduce the effects of power broadening and the ac Stark shift on the states. The $^{2}S_{1/2}$$\leftrightarrow$$^{2}P_{1/2}$ transition wavelengths are determined as the wavelength just before the fluorescence rapidly drops to background level, corresponding to heating of the ion, and the $^{2}D_{3/2}$$\leftrightarrow$$^{3}D[3/2]_{1/2}$ transition wavelengths are determined as the wavelengths at which maximum fluorescence is obtained. For the even isotopes, the external static magnetic field was removed while for $^{171}$Yb$^{+}$ a magnetic field of 0.5 mT was applied to ion to remove degeneracy within the hyperfine states, but keep the Zeeman shifts at a manageable level. Table \ref{wavelengths} shows the ionising, cooling and repumping wavelengths (vacuum) for the different Yb isotopes. Our measured 369.5 nm cooling wavelengths are in good agreement with those published by E. W. Streed et al. \cite{streed} who, using a Yb$^{+}$ hollow cathode lamp, observed a broadened Doppler absorption line at 369.525nm, with a width of $\approx$3 GHz, containing overlapping lines from the multiple ytterbium isotopes. However, our measurements are more precise. The intersection angle between the 398.9 nm laser beam and the average atomic velocity along the atomic beam result in a Doppler shift on the transition wavelength. In our setup the beam of neutral atoms and the 398.9 nm laser beam formed an angle of 63$^{o}$, while ionisation wavelengths corresponding to different angles can be found in reference \cite{Nizamani}. Due to infrequent population, precise transition wavelengths for the $^{2}$F$_{7/2}$ $\leftrightarrow$ $^{1}$D[5/2]$_{5/2}$ transition are difficult to obtain. Applying 120 Wcm$^{-2}$ to the ion, and using wavelengths of 638.618 nm for the even isotopes, and scanning between 638.610 nm and 638.616 nm for $^{171}$Yb$^{+}$ to account for hyperfine states \cite{roberts}, no obvious fluorescence interrupts were observed, indicating that these values are reasonably close to the exact transition wavelengths.

\begin{table}[!t]
\begin{center}

\begin{tabular}{c c c c}

    \hline
    \hline
    & $^{1}S_{0}$$\leftrightarrow$$^{1}P_{1}$ & $^{2}S_{1/2}$$\leftrightarrow$$^{2}P_{1/2}$ & $^{2}D_{3/2}$$\leftrightarrow$$^{3}D[3/2]_{1/2}$\\
    & transition & transition & transition\\
     & wavelength& wavelength& wavelength\\
    Isotope &(nm)&(nm)&(nm)\\ \hline
    $^{170}$Yb$^{+}$ & 398.91051(6) & 369.52364(6) & 935.19751(20) \\
    $^{171}$Yb$^{+}$ & 398.91070(6) & 369.52604(6) & 935.18768(20) \\
    $^{172}$Yb$^{+}$ & 398.91083(6) & 369.52435(6) & 935.18736(20) \\
    $^{174}$Yb$^{+}$ & 398.91114(6) & 369.52494(6) & 935.17976(20) \\
    $^{176}$Yb$^{+}$ & 398.91144(6) & 369.52550(6) & 935.17252(20) \\
    \hline
    \hline

\end{tabular}
\caption{Yb transition wavelengths (vacuum). The neutral atom $^{1}S_{0}$$\leftrightarrow$$^{1}P_{1}$ transition wavelengths apply to our setup where the neutral atomic beam and the laser beam make an angle of 63$^{o}$. The $^{2}S_{1/2}$$\leftrightarrow$$^{2}P_{1/2}$ and $^{2}D_{3/2}$$\leftrightarrow$$^{3}D[3/2]_{1/2}$ transition wavelengths were obtained by observing fluorescence from a trapped ion. The $^{171}$Yb$^{+}$ wavelengths correspond to the $^{2}S_{1/2}$(F = 1)$\leftrightarrow$$^{2}P_{1/2}$(F = 0) and $^{2}D_{3/2}$(F = 1)$\leftrightarrow$$^{3}D[3/2]_{1/2}$(F = 0) transitions respectively}
\label{wavelengths}
\end{center}
\end{table}

We have carried out a comprehensive error analysis and identify the the most significant source of error to be the absolute accuracy of the wavemeter. To eliminate any systematic offsets associated with the wavemeter it is calibrated using a 780 nm laser, which is locked to $<$1 MHz of the $^{87}$Rb D$_{2}$ line, and an addition He-Ne laser (calibrated to $<$1 MHz) is used to confirm the calibration. The wavemeter is calibrated before the measurements and the frequency standard was re-measured after the measurement to provide further confirmation of the calibration. The absolute accuracy of the wavemeter is specified to be 200 MHz below 370 nm and 60 MHz between 370 nm - 1100 nm, so the 369.5 nm wavelengths are inferred by measuring the wavelength of the 739.05 nm beam and doubling the frequency, resulting in an error of 120 MHz.

The intensity of the 369.5 nm beam is expected to broaden and AC stark shift the $^{2}S_{1/2}$$\leftrightarrow$$^{2}P_{1/2}$ transition by 40 MHz and 38 MHz respectively. As this transition is used for Doppler cooling, blue detuning the laser from the resonant transition frequency results in ion heating and an immediate drop in fluorescence, and therefore allows linewidth broadening to be neglected for these measurements. Fortunately, the resonance frequencies for this transition are determined by the immediate drop in fluorescence at the centre of resonance, and so is independent of linewidth broadening. Repeating the measurements at an intensity of 0.8 Wcm$^{-2}$, yields the same results. The intensity of the 935 nm beam broadens the transition to 30 MHz, and the AC stark shift is 17 MHz. In addition to the AC stark shift, the Zeeman shifts are also assessed. The magnetic field is removed for measurements of the even isotopes, leaving the ions exposed to low magnetic fields (such as that from the Earth), inducing a shift of only $\approx$1 MHz. The 0.5 mT external magnetic field applied when measuring $^{171}$Yb$^{+}$ is estimated to change the $\Delta$m$_{F} = \pm1$ transitions by $\approx \pm$10 MHz, while transition wavelengths between the magnetic field insensitive, $m_{F}=0$, states are not expected to change. Reducing the magnetic field to 0.25 mT reduced the $\Delta$m$_{F} = \pm1$ transitions by $\approx \pm$5 MHz, but due to the relatively large uncertainty associated with the wavemeter, no observable change in the resonance wavelengths were measured. Considering these effects the overall uncertainty for the $^{2}S_{1/2}$$\leftrightarrow$$^{2}P_{1/2}$ transition wavelengths is $\pm$126 MHz, while the uncertainty for $^{2}D_{3/2}$$\leftrightarrow$$^{3}D[3/2]_{1/2}$ transition wavelengths is $\pm$70 MHz.

\section{Heating Rate Measurement}
The ion heating rate was measured using a technique proposed by J. H. Wesenberg et al. \cite{NIST1}. A $^{174}$Yb$^{+}$ ion is trapped and allowed to heat by removing the 369.5 nm cooling beam, using an AOM (Isomet:1212-2-949, response time $\approx$ 100 ns), for periods of 1, 3, 5, and 7 seconds. The cooling laser is turned on again and the fluorescence is measured in 50 $\mu s$ bins for 4 ms. The particular technique requires the ion energy change to be dominated by heating along one of the principle axis of the trap. Assuming the heating rate has a $1/\omega^{2}$ dependance \cite{Turchette}, trap secular frequencies of $(\omega_{\text{x}},\omega_{\text{y}},\omega_{\text{z}})/2\pi$ = (2.069, 2.110, 0.178) $\pm$ 0.001 MHz result in dominant heating along the z-axis.

\begin{figure}[h]
\centering
\includegraphics[scale=0.27]{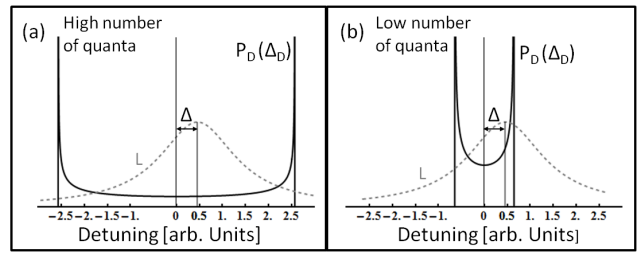}
\caption{Overlap of the probability density of the Doppler shift, $P_{\text{D}}(\Delta_{\text{D}})$ (solid line), with the transition linewidth $L$ (dashed line). (a) Diagram showing the case for a hot ion, with a maximum Doppler shift greater than the broadened transition linewidth and laser detuning, ${\Delta_{\text{max}}} \gg |L| + |{\Delta}|$. The overlap of both functions is low, leading to a low scatter rate. (b) Diagram showing the case for a cold ion with less energy. The peaks of $P_{\text{D}}(\Delta_{\text{D}})$ move closer together resulting in a stronger overlap and an increase in scatter rate.}
\label{Probab}
\end{figure}

Heating increases the ion's velocity, $v$, and the instantaneous Doppler shift during an oscillation is $\Delta_{\text{D}} = -kv$, where $k$ is the laser beam wavevector. We define $\Delta_{\text{max}}$ as the maximum instantaneous Doppler shift $\Delta_{\text{D}}$. The likelihood of the ion experiencing a specific instantaneous Doppler shift is described using a probability density, $P_{\text{D}}(\Delta_{\text{D}})$ \cite{NIST1}, and the overlap of $P_{\text{D}}(\Delta_{\text{D}})$ with the broadened FWHM transition linewidth, $L$, affects the scatter rate as shown in Fig. \ref{Probab}. Fig. \ref{Probab}(a) shows the situation for a hot ion with ${\Delta_{\text{max}}} \gg |L| + |{\Delta}|$, where $\Delta$ is the laser detuning from resonance. Here the Doppler shift probability density and transition linewidth poorly overlap, resulting in a low absorption/scatter rate. Fig. \ref{Probab}(b) shows the effects of a cold ion, where the probability density overlaps well with the transition linewidth producing an increased scatter rate.

The relative change of ion energy over one oscillation is small compared to the total energy change during the whole re-cooling experiment allowing the energy over an oscillation, and hence scatter rate, to be averaged. The average scatter rate $\left<dN/dt\right>$ over one oscillation is \cite{NIST1}

\begin{equation}
\left<\frac{dN}{dt}\right> = \int \frac{dN}{dt} P_{\text{D}}(\Delta_{\text{D}}) d\Delta_{\text{D}}
\end{equation}

where $dN/dt$ is the instantaneous scatter rate. In the regime of a hot ion, as depicted in Fig. \ref{Probab}(a), the probability density $P_{\text{D}}(\Delta_{\text{D}})$ is effectively uniform over the transition linewidth. The probability density can be factored out and allows the average scatter rate over one oscillation to be re-expressed in terms of the ion energy as

\begin{equation}
\label{dndt}
\left<\frac{dN}{dt}\right>(E(E_{0},t))= \frac{1}{\sqrt{E(E_{0},t)}}\frac{s L^{2}}{2\sqrt{\frac{2}{m}}k_{\text{z}}(1+s)^{3/2}}
\end{equation}

where $E(E_{0},t)$ is the ion energy at time $t$, $E_{0}$ is the ion energy just before recooling, $m$ is the ion mass, $k_{\text{z}}$ the z-component of the laser beam vector, $s$ the saturation intensity parameter, and $L$ the broadened transition linewidth. A 1D-Maxwell-Boltzmann distribution is used to describe the thermal distribution of the ion energy before recooling, $E_{0}$. The average scatter rate for an ion at time, $t$, is \cite{NIST1}

\begin{equation}
\label{HMscattertemp1}
\left<\frac{dN}{dt}\right>_{E_{0}} = \int_{\text{0}}^{\infty}  P_{\text{B}}(E_{\text{0}}) \left<\frac{dN}{dt}\right>(E(E_{\text{0}},t)) dE_{\text{0}}
\end{equation}

where $P_{\text{B}}(E_{0})$ is the 1D-Maxwell-Boltzmann distribution \cite{NIST1}. For the experiment, $\Delta$ = 6(2) MHz, s = 1.0(2), and the observed transition linewidth, $L$ = 40(5) MHz, is suspected to be due to power broadening and micromotion. The z component of the laser beam wave vector is $k_{\text{z}} = 0.45k$. To account for the thermal distribution of $E_{0}$ the recooling process was repeated 500 times and averaged. Fitting Eq. \ref{HMscattertemp1} to the average fluorescence curve determines the ion energy at the start of recooling.

\begin{figure}[h]
\centering
\includegraphics[scale=0.23]{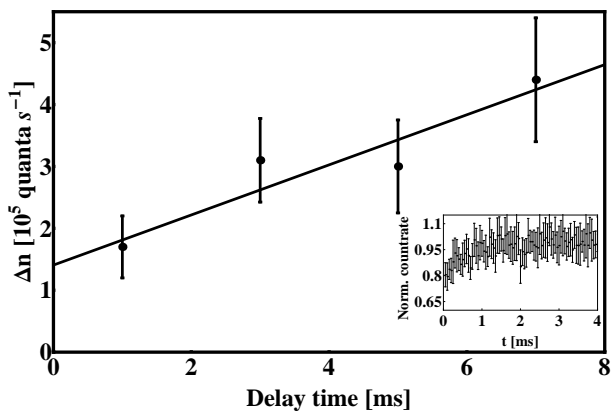}
\caption{Change in ion energy in terms of motional quanta, n, after heating periods of 1, 3, 5 and 7 seconds.  Each point results from 500 measurements. Inset: Change in fluorescence during the initial stages of recooling as a function of time.}
\label{HMcountrate}
\end{figure}

\begin{figure}[h]
\centering
\includegraphics[scale=0.48]{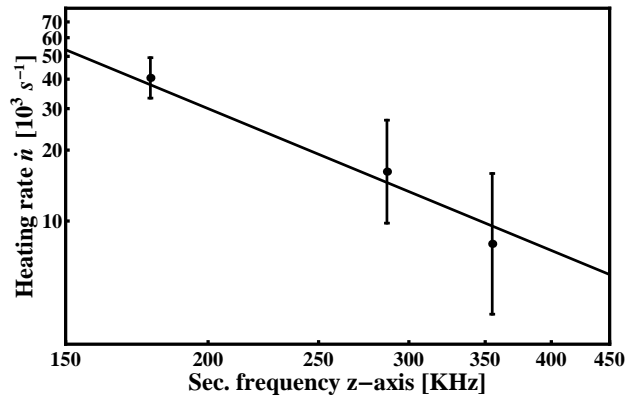}
\caption{Heating rate as a function of secular frequency. Heating measurements at secular frequencies of $\omega_{\text{z}}/2\pi$ = (178, 287, 355) $\pm$1 kHz, are consistent with a 1/$\omega^{2}$ dependance of the trap heating rate.}
\label{SE}
\end{figure}

The ion energy is converted into motional quanta ($\langle n \rangle = E/\hbar\omega_{\text{z}}$), and the change in motional quanta after different delay periods is shown in Fig. \ref{HMcountrate}. The inset shows the averaged fluorescence curve from 500 recooling cycles for a 5 second delay time. Repeating the experiment at secular frequencies of $\omega_{\text{z}} / 2\pi$ = (178, 287, 355) $\pm$1 kHz, produces the plot shown in Fig. \ref{SE} which is consistent with the expected $1/\omega^{2}$ dependance of $\dot{n}$. The cause of ion heating is the electric field noise density, $S_{E}(\omega_{\text{z}})$, which is related to the heating, $\langle \dot{n} \rangle$ \cite{Turchette,Deslauriers2} via

\begin{equation}
\label{fieldnoise}
\left< \dot{n} \right> = \frac{q^{2}}{4 m \hbar \omega_{\text{z}}} S_{\text{E}}(\omega_{\text{z}})
\end{equation}

where $q$ is the ion charge, $m$ the ion mass.

Assuming a $1/\omega^{2}$ dependency of $\dot{n}$ we calculate $S_{\text{E}}(1 \text{MHz})$ = 3.6(9)$\times10^{-11}$ V$^{2}$m$^{-2}$Hz$^{-1}$. Considering the ion electrode spacing of 310 $\mu m$ in our ion trap, the result is consistent with previously measured values of $S_{\text{E}}$ in other ion traps at room temperature \cite{Amini2}. This heating measurement, the first for a ytterbium ion in such a small ion trap structure, has mitigated a concern that electrodes coated with small amounts of ytterbium may lead to abnormally high heating rates. It is an encouraging result for the use of Yb$^{+}$ ions in quantum information processing.

\section{Conclusion}
We have shown how to create an experimental setup for the operation and development of advanced ion trap chips. In recent years the development of such ion trap designs has rapidly gained momentum and our setup allows for the simple and reliable operation and testing of such traps. This detailed description of the setup may therefore be very useful for the development of integrated ion chips. An ion trap was operated in this setup and we characterised its performance including a measurement of the motional heating rate. This measurement is consistent with previously reported rates of other ion species in different ion traps using established scaling laws \cite{Deslauriers2,Turchette,Labaziewicz,Labaziewicz2}. It further solidifies the suitability of the Yb$^{+}$ ion for quantum information processing as no abnormally high heating rates are observed in a relatively small ion trap. We also performed wavelength measurements for the $^{2}S_{1/2}$$\leftrightarrow$$^{2}P_{1/2}$ and $^{2}D_{3/2}$$\leftrightarrow$$^{3}D[3/2]_{1/2}$ transitions which are more precise than previous measurements. These measurements are particularly useful for groups setting up an ytterbium ion trap experiment as the availability of more precise frequency measurements significantly simplifies the initial trapping process.

\section{Acknowledgements}
We would like to thank undergraduate project students Daniel Brown, Nicholas Davies, Jessica Grove-Smith, Ben Pruess, Rajiv Ramasawmy, James Sayers, David Scrivener, Tim Short, and Philippa Young for their work in designing and building various components of the experimental setup. This work is supported by the UK Engineering and Physical Sciences Research Council (EP/E011136/1, EP/G007276/1), the European Commission's Sixth Framework Marie Curie International Reintegration Programme (MIRG-CT-2007-046432), the Nuffield Foundation and the University of Sussex.


\begin{thebibliography}{79}
\expandafter\ifx\csname natexlab\endcsname\relax\def\natexlab#1{#1}\fi
\expandafter\ifx\csname bibnamefont\endcsname\relax
  \def\bibnamefont#1{#1}\fi
\expandafter\ifx\csname bibfnamefont\endcsname\relax
  \def\bibfnamefont#1{#1}\fi
\expandafter\ifx\csname citenamefont\endcsname\relax
  \def\citenamefont#1{#1}\fi
\expandafter\ifx\csname url\endcsname\relax
  \def\url#1{\texttt{#1}}\fi
\expandafter\ifx\csname urlprefix\endcsname\relax\def\urlprefix{URL }\fi
\providecommand{\bibinfo}[2]{#2}
\providecommand{\eprint}[2][]{\url{#2}}

\bibitem[{\citenamefont{Cirac and Zoller}(1995)}]{zoller}
\bibinfo{author}{\bibfnamefont{J.~I.} \bibnamefont{Cirac}} \bibnamefont{and}
  \bibinfo{author}{\bibfnamefont{P.}~\bibnamefont{Zoller}},
  \bibinfo{journal}{Phys. Rev. Lett.} \textbf{\bibinfo{volume}{74}},
  \bibinfo{pages}{4091} (\bibinfo{year}{1995}).

\bibitem[{\citenamefont{Wineland
  et~al.}(1998{\natexlab{a}})\citenamefont{Wineland, Monroe, Itano, Leibfried,
  King, and Meekhof}}]{wineland}
\bibinfo{author}{\bibfnamefont{D.~J.} \bibnamefont{Wineland}},
  \bibinfo{author}{\bibfnamefont{C.}~\bibnamefont{Monroe}},
  \bibinfo{author}{\bibfnamefont{W.~M.} \bibnamefont{Itano}},
  \bibinfo{author}{\bibfnamefont{D.}~\bibnamefont{Leibfried}},
  \bibinfo{author}{\bibfnamefont{B.~E.} \bibnamefont{King}}, \bibnamefont{and}
  \bibinfo{author}{\bibfnamefont{D.~M.} \bibnamefont{Meekhof}},
  \bibinfo{journal}{J. Res. Natl. Inst. Stand. Technol.}
  \textbf{\bibinfo{volume}{103}}, \bibinfo{pages}{259}
  (\bibinfo{year}{1998}{\natexlab{a}}).

\bibitem[{\citenamefont{H\"affner et~al.}(2008)\citenamefont{H\"affner, Roos,
  and Blatt}}]{haffner}
\bibinfo{author}{\bibfnamefont{H.}~\bibnamefont{H\"affner}},
  \bibinfo{author}{\bibfnamefont{C.~F.} \bibnamefont{Roos}}, \bibnamefont{and}
  \bibinfo{author}{\bibfnamefont{R.}~\bibnamefont{Blatt}},
  \bibinfo{journal}{Phys. Rep.} \textbf{\bibinfo{volume}{469}},
  \bibinfo{pages}{155} (\bibinfo{year}{2008}).

\bibitem[{\citenamefont{Pons et~al.}(2007)\citenamefont{Pons, Ahufinger,
  Wunderlich, Sanpera, Braungardt, Sen(De), Sen, and Lewenstein}}]{pons}
\bibinfo{author}{\bibfnamefont{M.}~\bibnamefont{Pons}},
  \bibinfo{author}{\bibfnamefont{V.}~\bibnamefont{Ahufinger}},
  \bibinfo{author}{\bibfnamefont{C.}~\bibnamefont{Wunderlich}},
  \bibinfo{author}{\bibfnamefont{A.}~\bibnamefont{Sanpera}},
  \bibinfo{author}{\bibfnamefont{S.}~\bibnamefont{Braungardt}},
  \bibinfo{author}{\bibfnamefont{A.}~\bibnamefont{Sen(De)}},
  \bibinfo{author}{\bibfnamefont{U.}~\bibnamefont{Sen}}, \bibnamefont{and}
  \bibinfo{author}{\bibfnamefont{M.}~\bibnamefont{Lewenstein}},
  \bibinfo{journal}{Phys. Rev. Lett.} \textbf{\bibinfo{volume}{98}},
  \bibinfo{pages}{023003} (\bibinfo{year}{2007}).

\bibitem[{\citenamefont{Friedenauer et~al.}(2008)\citenamefont{Friedenauer,
  Schmitz, Glueckert, Porras, and Schaetz}}]{friedenaueri}
\bibinfo{author}{\bibfnamefont{A.}~\bibnamefont{Friedenauer}},
  \bibinfo{author}{\bibfnamefont{H.}~\bibnamefont{Schmitz}},
  \bibinfo{author}{\bibfnamefont{J.~T.} \bibnamefont{Glueckert}},
  \bibinfo{author}{\bibfnamefont{D.}~\bibnamefont{Porras}}, \bibnamefont{and}
  \bibinfo{author}{\bibfnamefont{T.}~\bibnamefont{Schaetz}},
  \bibinfo{journal}{Nature Phys.} \textbf{\bibinfo{volume}{4}},
  \bibinfo{pages}{757} (\bibinfo{year}{2008}).

\bibitem[{\citenamefont{Johanning et~al.}(2009)\citenamefont{Johanning,
  Var\'on, and Wunderlich}}]{Johanning}
\bibinfo{author}{\bibfnamefont{M.}~\bibnamefont{Johanning}},
  \bibinfo{author}{\bibfnamefont{A.~F.} \bibnamefont{Var\'on}},
  \bibnamefont{and}
  \bibinfo{author}{\bibfnamefont{C.}~\bibnamefont{Wunderlich}},
  \bibinfo{journal}{J. Phys. B} \textbf{\bibinfo{volume}{42}},
  \bibinfo{pages}{154009} (\bibinfo{year}{2009}).

\bibitem[{\citenamefont{Clark et~al.}(2009)\citenamefont{Clark, Lin, Brown, and
  Chuang}}]{clarki}
\bibinfo{author}{\bibfnamefont{R.~J.} \bibnamefont{Clark}},
  \bibinfo{author}{\bibfnamefont{T.}~\bibnamefont{Lin}},
  \bibinfo{author}{\bibfnamefont{K.~R.} \bibnamefont{Brown}}, \bibnamefont{and}
  \bibinfo{author}{\bibfnamefont{I.~L.} \bibnamefont{Chuang}},
  \bibinfo{journal}{J. Appl. Phys.} \textbf{\bibinfo{volume}{105}},
  \bibinfo{pages}{013114} (\bibinfo{year}{2009}).

\bibitem[{\citenamefont{Kim et~al.}(2010)\citenamefont{Kim, Chang, Korenblit,
  Islam, Edwards, Freericks, Lin, Duan, and Monroe}}]{kim}
\bibinfo{author}{\bibfnamefont{K.}~\bibnamefont{Kim}},
  \bibinfo{author}{\bibfnamefont{M.-S.} \bibnamefont{Chang}},
  \bibinfo{author}{\bibfnamefont{S.}~\bibnamefont{Korenblit}},
  \bibinfo{author}{\bibfnamefont{R.}~\bibnamefont{Islam}},
  \bibinfo{author}{\bibfnamefont{E.~E.} \bibnamefont{Edwards}},
  \bibinfo{author}{\bibfnamefont{J.~K.} \bibnamefont{Freericks}},
  \bibinfo{author}{\bibfnamefont{G.-D.} \bibnamefont{Lin}},
  \bibinfo{author}{\bibfnamefont{L.-M.} \bibnamefont{Duan}}, \bibnamefont{and}
  \bibinfo{author}{\bibfnamefont{C.}~\bibnamefont{Monroe}},
  \bibinfo{journal}{Nature (London)} \textbf{\bibinfo{volume}{465}}, \bibinfo{pages}{590} (\bibinfo{year}{2010}).

\bibitem[{\citenamefont{Udem et~al.}(2001)\citenamefont{Udem, Diddams, Vogel,
  Oates, Curtis, Lee, Itano, Drullinger, Bergquist, and Hollberg}}]{udem}
\bibinfo{author}{\bibfnamefont{Th.}~\bibnamefont{Udem}},
  \bibinfo{author}{\bibfnamefont{S.~A.} \bibnamefont{Diddams}},
  \bibinfo{author}{\bibfnamefont{K.~R.} \bibnamefont{Vogel}},
  \bibinfo{author}{\bibfnamefont{C.~W.} \bibnamefont{Oates}},
  \bibinfo{author}{\bibfnamefont{E.~A.} \bibnamefont{Curtis}},
  \bibinfo{author}{\bibfnamefont{W.~D.} \bibnamefont{Lee}},
  \bibinfo{author}{\bibfnamefont{W.~M.} \bibnamefont{Itano}},
  \bibinfo{author}{\bibfnamefont{R.~E.} \bibnamefont{Drullinger}},
  \bibinfo{author}{\bibfnamefont{J.~C.} \bibnamefont{Bergquist}},
  \bibnamefont{and} \bibinfo{author}{\bibfnamefont{L.}~\bibnamefont{Hollberg}},
  \bibinfo{journal}{Phys. Rev. Lett.} \textbf{\bibinfo{volume}{86}},
  \bibinfo{pages}{4996} (\bibinfo{year}{2001}).

\bibitem[{\citenamefont{Webster et~al.}(2002)\citenamefont{Webster, Taylor,
  Roberts, Barwood, and Gill}}]{Webster}
\bibinfo{author}{\bibfnamefont{S.~A.} \bibnamefont{Webster}},
  \bibinfo{author}{\bibfnamefont{P.}~\bibnamefont{Taylor}},
  \bibinfo{author}{\bibfnamefont{M.}~\bibnamefont{Roberts}},
  \bibinfo{author}{\bibfnamefont{G.~P.} \bibnamefont{Barwood}},
  \bibnamefont{and} \bibinfo{author}{\bibfnamefont{P.}~\bibnamefont{Gill}},
  \bibinfo{journal}{Phys. Rev. A} \textbf{\bibinfo{volume}{65}},
  \bibinfo{pages}{052501} (\bibinfo{year}{2002}).

\bibitem[{\citenamefont{Tamm et~al.}(2009)\citenamefont{Tamm, Weyers,
  Lipphardt, and Peik}}]{Tamm}
\bibinfo{author}{\bibfnamefont{C.}~\bibnamefont{Tamm}},
  \bibinfo{author}{\bibfnamefont{S.}~\bibnamefont{Weyers}},
  \bibinfo{author}{\bibfnamefont{B.}~\bibnamefont{Lipphardt}},
  \bibnamefont{and} \bibinfo{author}{\bibfnamefont{E.}~\bibnamefont{Peik}},
  \bibinfo{journal}{Phys. Rev. A} \textbf{\bibinfo{volume}{80}},
  \bibinfo{pages}{043403} (\bibinfo{year}{2009}).

\bibitem[{\citenamefont{Chwalla et~al.}(2009)\citenamefont{Chwalla, Benhelm,
  Kim, Kirchmair, Monz, Riebe, Schindler, Villar, H\"ansel, Roos, Blatt, Abgrall, Santarwlli, Rovera, and Laurent}}]{chwalla}
\bibinfo{author}{\bibfnamefont{M.}~\bibnamefont{Chwalla}},
  \bibinfo{author}{\bibfnamefont{J.}~\bibnamefont{Benhelm}},
  \bibinfo{author}{\bibfnamefont{K.}~\bibnamefont{Kim}},
  \bibinfo{author}{\bibfnamefont{G.}~\bibnamefont{Kirchmair}},
  \bibinfo{author}{\bibfnamefont{T.}~\bibnamefont{Monz}},
  \bibinfo{author}{\bibfnamefont{M.}~\bibnamefont{Riebe}},
  \bibinfo{author}{\bibfnamefont{P.}~\bibnamefont{Schindler}},
  \bibinfo{author}{\bibfnamefont{A.~S.} \bibnamefont{Villar}},
  \bibinfo{author}{\bibfnamefont{W.}~\bibnamefont{H\"ansel}},
  \bibinfo{author}{\bibfnamefont{C.~F.} \bibnamefont{Roos}},
  \bibinfo{author}{\bibfnamefont{R.} \bibnamefont{Blatt}},
  \bibinfo{author}{\bibfnamefont{M.} \bibnamefont{Abgrall}},
  \bibinfo{author}{\bibfnamefont{G.} \bibnamefont{Santarelli}},
  \bibinfo{author}{\bibfnamefont{G.~D.} \bibnamefont{Rovera}},
    \bibnamefont{and}
   \bibinfo{author}{\bibfnamefont{P.~H.} \bibnamefont{Laurent}}, \bibinfo{journal}{Phys. Rev. Lett.}
  \textbf{\bibinfo{volume}{102}}, \bibinfo{pages}{023002}
  (\bibinfo{year}{2009}).

\bibitem[{\citenamefont{Schaetz et~al.}(2005)\citenamefont{Schaetz, Barrett,
  Leibfried, Britton, Chiaverini, Itano, Jost, Knill, Langer, and
  Wineland}}]{schaetzi}
\bibinfo{author}{\bibfnamefont{T.}~\bibnamefont{Schaetz}},
  \bibinfo{author}{\bibfnamefont{M.~D.} \bibnamefont{Barrett}},
  \bibinfo{author}{\bibfnamefont{D.}~\bibnamefont{Leibfried}},
  \bibinfo{author}{\bibfnamefont{J.}~\bibnamefont{Britton}},
  \bibinfo{author}{\bibfnamefont{J.}~\bibnamefont{Chiaverini}},
  \bibinfo{author}{\bibfnamefont{W.~M.} \bibnamefont{Itano}},
  \bibinfo{author}{\bibfnamefont{J.~D.} \bibnamefont{Jost}},
  \bibinfo{author}{\bibfnamefont{E.}~\bibnamefont{Knill}},
  \bibinfo{author}{\bibfnamefont{C.}~\bibnamefont{Langer}}, \bibnamefont{and}
  \bibinfo{author}{\bibfnamefont{D.~J.} \bibnamefont{Wineland}},
  \bibinfo{journal}{Phys. Rev. Lett.} \textbf{\bibinfo{volume}{94}},
  \bibinfo{pages}{010501} (\bibinfo{year}{2005}).

\bibitem[{\citenamefont{Acton et~al.}(2006)\citenamefont{Acton, Brickman,
  Haljan, Lee, Deslauriers, and Monroe}}]{actoni}
\bibinfo{author}{\bibfnamefont{M.}~\bibnamefont{Acton}},
  \bibinfo{author}{\bibfnamefont{K.-A.}~\bibnamefont{Brickman}},
  \bibinfo{author}{\bibfnamefont{P.~C.}~\bibnamefont{Haljan}},
  \bibinfo{author}{\bibfnamefont{P.~J.}~\bibnamefont{Lee}},
  \bibinfo{author}{\bibfnamefont{L.}~\bibnamefont{Deslauriers}},
  \bibnamefont{and} \bibinfo{author}{\bibfnamefont{C.}~\bibnamefont{Monroe}},
  \bibinfo{journal}{Quantum Inf. Comput}. \textbf{\bibinfo{volume}{6}},
  \bibinfo{pages}{465} (\bibinfo{year}{2006}).

\bibitem[{\citenamefont{Wunderlich et~al.}(2007)\citenamefont{Wunderlich,
  Th.~Hannemann, H\"affner, Roos, H\"ansel, Blatt, and
  Schmidt-Kaler}}]{Wunderlich}
\bibinfo{author}{\bibfnamefont{C.}~\bibnamefont{Wunderlich}},
  \bibinfo{author}{\bibfnamefont{T.~K.} \bibnamefont{Th.~Hannemann}},
  \bibinfo{author}{\bibfnamefont{H.}~\bibnamefont{H\"affner}},
  \bibinfo{author}{\bibfnamefont{C.}~\bibnamefont{Roos}},
  \bibinfo{author}{\bibfnamefont{W.}~\bibnamefont{H\"ansel}},
  \bibinfo{author}{\bibfnamefont{R.}~\bibnamefont{Blatt}}, \bibnamefont{and}
  \bibinfo{author}{\bibfnamefont{F.}~\bibnamefont{Schmidt-Kaler}},
  \bibinfo{journal}{J. Mod. Opt.} \textbf{\bibinfo{volume}{54}},
  \bibinfo{pages}{1541} (\bibinfo{year}{2007}).

\bibitem[{\citenamefont{Myerson et~al.}(2008)\citenamefont{Myerson, Szwer,
  Webster, Allcock, Curtis, Imreh, Sherman, Stacey, Steane, and
  Lucas}}]{myersoni}
\bibinfo{author}{\bibfnamefont{A.~H.} \bibnamefont{Myerson}},
  \bibinfo{author}{\bibfnamefont{D.~J.} \bibnamefont{Szwer}},
  \bibinfo{author}{\bibfnamefont{S.~C.} \bibnamefont{Webster}},
  \bibinfo{author}{\bibfnamefont{D.~T.~C.} \bibnamefont{Allcock}},
  \bibinfo{author}{\bibfnamefont{M.~J.} \bibnamefont{Curtis}},
  \bibinfo{author}{\bibfnamefont{G.}~\bibnamefont{Imreh}},
  \bibinfo{author}{\bibfnamefont{J.~A.} \bibnamefont{Sherman}},
  \bibinfo{author}{\bibfnamefont{D.~N.} \bibnamefont{Stacey}},
  \bibinfo{author}{\bibfnamefont{A.~M.} \bibnamefont{Steane}},
  \bibnamefont{and} \bibinfo{author}{\bibfnamefont{D.~M.} \bibnamefont{Lucas}},
  \bibinfo{journal}{Phys. Rev. Lett.} \textbf{\bibinfo{volume}{100}},
  \bibinfo{pages}{200502} (\bibinfo{year}{2008}).

\bibitem[{\citenamefont{Burrell et~al.}(2010)\citenamefont{Burrell, Szwer,
  Webster, and Lucas}}]{burrell2}
\bibinfo{author}{\bibfnamefont{A.~H.} \bibnamefont{Burrell}},
  \bibinfo{author}{\bibfnamefont{D.~J.} \bibnamefont{Szwer}},
  \bibinfo{author}{\bibfnamefont{S.~C.} \bibnamefont{Webster}},
  \bibnamefont{and} \bibinfo{author}{\bibfnamefont{D.~M.} \bibnamefont{Lucas}},
  \bibinfo{journal}{Phys. Rev. A} \textbf{\bibinfo{volume}{81}},
  \bibinfo{pages}{040302} (\bibinfo{year}{2010}).

\bibitem[{\citenamefont{Sackett et~al.}(2000)\citenamefont{Sackett, Kielpinski,
  King, Langer, Meyer, Myatt, Rowe, Turchette, Itano, and Wineland}}]{sacketti}
\bibinfo{author}{\bibfnamefont{C.~A.}~\bibnamefont{Sackett}},
  \bibinfo{author}{\bibfnamefont{D.}~\bibnamefont{Kielpinski}},
  \bibinfo{author}{\bibfnamefont{B.~E.}~\bibnamefont{King}},
  \bibinfo{author}{\bibfnamefont{C.}~\bibnamefont{Langer}},
  \bibinfo{author}{\bibfnamefont{V.}~\bibnamefont{Meyer}},
  \bibinfo{author}{\bibfnamefont{C.~J.}~\bibnamefont{Myatt}},
  \bibinfo{author}{\bibfnamefont{M.}~\bibnamefont{Rowe}},
  \bibinfo{author}{\bibfnamefont{Q.~A.}~\bibnamefont{Turchette}},
  \bibinfo{author}{\bibfnamefont{W.~M.}~\bibnamefont{Itano}},
    \bibnamefont{and}
  \bibinfo{author}{\bibfnamefont{D.~J.}~\bibnamefont{Wineland}}, \bibinfo{journal}{Nature (London)}
  \textbf{\bibinfo{volume}{404}}, \bibinfo{pages}{256} (\bibinfo{year}{2000}).

\bibitem[{\citenamefont{Leibfried et~al.}(2003)\citenamefont{Leibfried,
  DeMarco, Meyer, Lucas, Barrett, Britton, Itano, Jelenkovi\'{c}, Langer, Rosenband, and Wineland}}]{leibfriedai}
\bibinfo{author}{\bibfnamefont{D.}~\bibnamefont{Leibfried}},
  \bibinfo{author}{\bibfnamefont{B.}~\bibnamefont{DeMarco}},
  \bibinfo{author}{\bibfnamefont{V.}~\bibnamefont{Meyer}},
  \bibinfo{author}{\bibfnamefont{D.}~\bibnamefont{Lucas}},
  \bibinfo{author}{\bibfnamefont{M.}~\bibnamefont{Barrett}},
  \bibinfo{author}{\bibfnamefont{J.}~\bibnamefont{Britton}},
  \bibinfo{author}{\bibfnamefont{W.~M.} \bibnamefont{Itano}},
  \bibinfo{author}{\bibfnamefont{B.}~\bibnamefont{Jelenkovi\'{c}}},
  \bibinfo{author}{\bibfnamefont{C.}~\bibnamefont{Langer}},
  \bibinfo{author}{\bibfnamefont{T.}~\bibnamefont{Rosenband}},
      \bibnamefont{and}
  \bibinfo{author}{\bibfnamefont{D.~J.}~\bibnamefont{Wineland}},
 \bibinfo{journal}{Nature} \textbf{\bibinfo{volume}{422}}, \bibinfo{pages}{412} (\bibinfo{year}{2003}).

\bibitem[{\citenamefont{Schmidt-Kaler et~al.}(2003)\citenamefont{Schmidt-Kaler,
  H\"affner, Riebe, Gulde, Lancaster, Deuschle, Becher, Roos, Eschner, and
  Blatt}}]{schmidti}
\bibinfo{author}{\bibfnamefont{F.}~\bibnamefont{Schmidt-Kaler}},
  \bibinfo{author}{\bibfnamefont{H.}~\bibnamefont{H\"affner}},
  \bibinfo{author}{\bibfnamefont{M.}~\bibnamefont{Riebe}},
  \bibinfo{author}{\bibfnamefont{S.}~\bibnamefont{Gulde}},
  \bibinfo{author}{\bibfnamefont{G.~P.~T.} \bibnamefont{Lancaster}},
  \bibinfo{author}{\bibfnamefont{T.}~\bibnamefont{Deuschle}},
  \bibinfo{author}{\bibfnamefont{C.}~\bibnamefont{Becher}},
  \bibinfo{author}{\bibfnamefont{C.~F.} \bibnamefont{Roos}},
  \bibinfo{author}{\bibfnamefont{J.}~\bibnamefont{Eschner}}, \bibnamefont{and}
  \bibinfo{author}{\bibfnamefont{R.}~\bibnamefont{Blatt}},
  \bibinfo{journal}{Nature (London)} \textbf{\bibinfo{volume}{422}}, \bibinfo{pages}{408} (\bibinfo{year}{2003}).

\bibitem[{\citenamefont{Chiaverini et~al.}(2004)\citenamefont{Chiaverini,
  Leibfried, Schaetz, Barrett, Blakestad, Britton, Itano, Jost, Knill, Langer, Ozeri, and Wineland}}]{chiaverinii}
\bibinfo{author}{\bibfnamefont{J.}~\bibnamefont{Chiaverini}},
  \bibinfo{author}{\bibfnamefont{D.}~\bibnamefont{Leibfried}},
  \bibinfo{author}{\bibfnamefont{T.}~\bibnamefont{Schaetz}},
  \bibinfo{author}{\bibfnamefont{M.~D.}~\bibnamefont{Barrett}},
  \bibinfo{author}{\bibfnamefont{R.~B.}~\bibnamefont{Blakestad}},
  \bibinfo{author}{\bibfnamefont{J.}~\bibnamefont{Britton}},
  \bibinfo{author}{\bibfnamefont{W.~M.}~\bibnamefont{Itano}},
  \bibinfo{author}{\bibfnamefont{J.~D.}~\bibnamefont{Jost}},
  \bibinfo{author}{\bibfnamefont{E.}~\bibnamefont{Knill}},
  \bibinfo{author}{\bibfnamefont{C.}~\bibnamefont{Langer}},
  \bibinfo{author}{\bibfnamefont{R.}~\bibnamefont{Ozeri}},
      \bibnamefont{and}
  \bibinfo{author}{\bibfnamefont{D.~J.}~\bibnamefont{Wineland}},
  \bibinfo{journal}{Nature (London)}
  \textbf{\bibinfo{volume}{432}}, \bibinfo{pages}{602}
  (\bibinfo{year}{2004}).

\bibitem[{\citenamefont{Brickman et~al.}(2005)\citenamefont{Brickman, Haljan,
  Lee, Acton, Deslauriers, and Monroe}}]{brickman}
\bibinfo{author}{\bibfnamefont{K.-A.} \bibnamefont{Brickman}},
  \bibinfo{author}{\bibfnamefont{P.~C.} \bibnamefont{Haljan}},
  \bibinfo{author}{\bibfnamefont{P.~J.} \bibnamefont{Lee}},
  \bibinfo{author}{\bibfnamefont{M.}~\bibnamefont{Acton}},
  \bibinfo{author}{\bibfnamefont{L.}~\bibnamefont{Deslauriers}},
  \bibnamefont{and} \bibinfo{author}{\bibfnamefont{C.}~\bibnamefont{Monroe}},
  \bibinfo{journal}{Phys. Rev. A} \textbf{\bibinfo{volume}{72}},
  \bibinfo{pages}{050306(R)} (\bibinfo{year}{2005}).

\bibitem[{\citenamefont{Home et~al.}(2006)\citenamefont{Home, McDonnell, Lucas,
  Imreh, Keitch, Szwer, Thomas, Webster, Stacey, and Steane}}]{home}
\bibinfo{author}{\bibfnamefont{J.~P.} \bibnamefont{Home}},
  \bibinfo{author}{\bibfnamefont{M.~J.} \bibnamefont{McDonnell}},
  \bibinfo{author}{\bibfnamefont{D.~M.} \bibnamefont{Lucas}},
  \bibinfo{author}{\bibfnamefont{G.}~\bibnamefont{Imreh}},
  \bibinfo{author}{\bibfnamefont{B.~C.} \bibnamefont{Keitch}},
  \bibinfo{author}{\bibfnamefont{D.~J.} \bibnamefont{Szwer}},
  \bibinfo{author}{\bibfnamefont{N.~R.} \bibnamefont{Thomas}},
  \bibinfo{author}{\bibfnamefont{S.~C.} \bibnamefont{Webster}},
  \bibinfo{author}{\bibfnamefont{D.~N.} \bibnamefont{Stacey}},
  \bibnamefont{and} \bibinfo{author}{\bibfnamefont{A.~M.}
  \bibnamefont{Steane}}, \bibinfo{journal}{New J. Phys.}
  \textbf{\bibinfo{volume}{8}}, \bibinfo{pages}{188} (\bibinfo{year}{2006}).

\bibitem[{\citenamefont{Ospelkaus et~al.}(2008)\citenamefont{Ospelkaus, Langer, Amini, Brown, Leibfried, and Wineland}}]{Ospelkaus}
\bibinfo{author}{\bibfnamefont{C.}~\bibnamefont{Ospelkaus}},
\bibinfo{author}{\bibfnamefont{C.~E.}~\bibnamefont{Langer}},
  \bibinfo{author}{\bibfnamefont{J.~M.} \bibnamefont{Amini}},
  \bibinfo{author}{\bibfnamefont{K.~R.} \bibnamefont{Brown}},
  \bibinfo{author}{\bibfnamefont{D.}~\bibnamefont{Leibfried}},
  \bibnamefont{and} \bibinfo{author}{\bibfnamefont{D.~J.}
  \bibnamefont{Wineland}}, \bibinfo{journal}{Phys. Rev. Lett.}
  \textbf{\bibinfo{volume}{101}}, \bibinfo{pages}{090502}
  (\bibinfo{year}{2008}).

\bibitem[{\citenamefont{Monz et~al.}(2009)\citenamefont{Monz, Kim, H\"ansel,
  Riebe, Villar, Schindler, Chwalla, Hennrich, and Blatt}}]{monz}
\bibinfo{author}{\bibfnamefont{T.}~\bibnamefont{Monz}},
  \bibinfo{author}{\bibfnamefont{K.}~\bibnamefont{Kim}},
  \bibinfo{author}{\bibfnamefont{W.}~\bibnamefont{H\"ansel}},
  \bibinfo{author}{\bibfnamefont{M.}~\bibnamefont{Riebe}},
  \bibinfo{author}{\bibfnamefont{A.~S.} \bibnamefont{Villar}},
  \bibinfo{author}{\bibfnamefont{P.}~\bibnamefont{Schindler}},
  \bibinfo{author}{\bibfnamefont{M.}~\bibnamefont{Chwalla}},
  \bibinfo{author}{\bibfnamefont{M.}~\bibnamefont{Hennrich}}, \bibnamefont{and}
  \bibinfo{author}{\bibfnamefont{R.}~\bibnamefont{Blatt}},
  \bibinfo{journal}{Phys. Rev. Lett.} \textbf{\bibinfo{volume}{102}},
  \bibinfo{pages}{040501} (\bibinfo{year}{2009}).

\bibitem[{\citenamefont{Wang et~al.}(2010)\citenamefont{Wang, Labaziewicz, Ge,
  Shewmon, and Chuang}}]{wang}
\bibinfo{author}{\bibfnamefont{S.~X.} \bibnamefont{Wang}},
  \bibinfo{author}{\bibfnamefont{J.}~\bibnamefont{Labaziewicz}},
  \bibinfo{author}{\bibfnamefont{Y.}~\bibnamefont{Ge}},
  \bibinfo{author}{\bibfnamefont{R.}~\bibnamefont{Shewmon}}, \bibnamefont{and}
  \bibinfo{author}{\bibfnamefont{I.~L.} \bibnamefont{Chuang}},
  \bibinfo{journal}{Phys. Rev. A} \textbf{\bibinfo{volume}{81}},
  \bibinfo{pages}{062332} (\bibinfo{year}{2010}).

\bibitem[{\citenamefont{Timoney et~al.}(2008)\citenamefont{Timoney, Elman,
  Glaser, Weiss, Johanning, Neuhauser, and Wunderlich}}]{timoney}
\bibinfo{author}{\bibfnamefont{N.}~\bibnamefont{Timoney}},
  \bibinfo{author}{\bibfnamefont{V.}~\bibnamefont{Elman}},
  \bibinfo{author}{\bibfnamefont{S.}~\bibnamefont{Glaser}},
  \bibinfo{author}{\bibfnamefont{C.}~\bibnamefont{Weiss}},
  \bibinfo{author}{\bibfnamefont{M.}~\bibnamefont{Johanning}},
  \bibinfo{author}{\bibfnamefont{W.}~\bibnamefont{Neuhauser}},
  \bibnamefont{and}
  \bibinfo{author}{\bibfnamefont{C.}~\bibnamefont{Wunderlich}},
  \bibinfo{journal}{Phys. Rev. A} \textbf{\bibinfo{volume}{77}},
  \bibinfo{pages}{052334} (\bibinfo{year}{2008}).

\bibitem[{\citenamefont{Rowe et~al.}(2002)\citenamefont{Rowe, Ben-Kish,
  DeMarco, Leibfried, Meyer, Beall, Britton, Hughes, Itano, Jelenkovi\'{c}
  Rosenbrnd, and Wineland}}]{rowe}
\bibinfo{author}{\bibfnamefont{M.~A.} \bibnamefont{Rowe}},
  \bibinfo{author}{\bibfnamefont{A.}~\bibnamefont{Ben-Kish}},
  \bibinfo{author}{\bibfnamefont{B.}~\bibnamefont{DeMarco}},
  \bibinfo{author}{\bibfnamefont{D.}~\bibnamefont{Leibfried}},
  \bibinfo{author}{\bibfnamefont{V.}~\bibnamefont{Meyer}},
  \bibinfo{author}{\bibfnamefont{J.}~\bibnamefont{Beall}},
  \bibinfo{author}{\bibfnamefont{J.}~\bibnamefont{Britton}},
  \bibinfo{author}{\bibfnamefont{J.}~\bibnamefont{Hughes}},
  \bibinfo{author}{\bibfnamefont{W.~M.} \bibnamefont{Itano}},
  \bibinfo{author}{\bibfnamefont{B.}~\bibnamefont{Jelenkovi\'{c}}},
    \bibinfo{author}{\bibfnamefont{T.} \bibnamefont{Rosenbrnd}},
        \bibnamefont{and}
  \bibinfo{author}{\bibfnamefont{D.~J.}~\bibnamefont{Wineland}}, \bibinfo{journal}{Quantum Inf. Comput.}
  \textbf{\bibinfo{volume}{2}}, \bibinfo{pages}{257} (\bibinfo{year}{2002}).

\bibitem[{\citenamefont{Hensinger et~al.}(2006)\citenamefont{Hensinger,
  Olmschenk, Stick, Hucul, Yeo, Acton, Deslauriers, Monroe, and
  Rabchuk}}]{hensinger}
\bibinfo{author}{\bibfnamefont{W.~K.} \bibnamefont{Hensinger}},
  \bibinfo{author}{\bibfnamefont{S.}~\bibnamefont{Olmschenk}},
  \bibinfo{author}{\bibfnamefont{D.}~\bibnamefont{Stick}},
  \bibinfo{author}{\bibfnamefont{D.}~\bibnamefont{Hucul}},
  \bibinfo{author}{\bibfnamefont{M.}~\bibnamefont{Yeo}},
  \bibinfo{author}{\bibfnamefont{M.}~\bibnamefont{Acton}},
  \bibinfo{author}{\bibfnamefont{L.}~\bibnamefont{Deslauriers}},
  \bibinfo{author}{\bibfnamefont{C.}~\bibnamefont{Monroe}}, \bibnamefont{and}
  \bibinfo{author}{\bibfnamefont{J.}~\bibnamefont{Rabchuk}},
  \bibinfo{journal}{Appl. Phys. Lett.} \textbf{\bibinfo{volume}{88}},
  \bibinfo{pages}{034101} (\bibinfo{year}{2006}).

\bibitem[{\citenamefont{Schulz et~al.}(2006)\citenamefont{Schulz, Poschinger,
  Singer, and Schmidt-Kaler}}]{schulz1}
\bibinfo{author}{\bibfnamefont{S.}~\bibnamefont{Schulz}},
  \bibinfo{author}{\bibfnamefont{U.}~\bibnamefont{Poschinger}},
  \bibinfo{author}{\bibfnamefont{K.}~\bibnamefont{Singer}}, \bibnamefont{and}
  \bibinfo{author}{\bibfnamefont{F.}~\bibnamefont{Schmidt-Kaler}},
  \bibinfo{journal}{Fortschr. Phys.} \textbf{\bibinfo{volume}{54}},
  \bibinfo{pages}{648} (\bibinfo{year}{2006}).

\bibitem[{\citenamefont{Reichle et~al.}(2006)\citenamefont{Reichle, Leibfried,
  Blakestad, Britton, Jost, Knill, Langer, Ozeri, Seidelin, and
  J.Wineland}}]{reichle}
\bibinfo{author}{\bibfnamefont{R.}~\bibnamefont{Reichle}},
  \bibinfo{author}{\bibfnamefont{D.}~\bibnamefont{Leibfried}},
  \bibinfo{author}{\bibfnamefont{R.}~\bibnamefont{Blakestad}},
  \bibinfo{author}{\bibfnamefont{J.}~\bibnamefont{Britton}},
  \bibinfo{author}{\bibfnamefont{J.}~\bibnamefont{Jost}},
  \bibinfo{author}{\bibfnamefont{E.}~\bibnamefont{Knill}},
  \bibinfo{author}{\bibfnamefont{C.}~\bibnamefont{Langer}},
  \bibinfo{author}{\bibfnamefont{R.}~\bibnamefont{Ozeri}},
  \bibinfo{author}{\bibfnamefont{S.}~\bibnamefont{Seidelin}}, \bibnamefont{and}
  \bibinfo{author}{\bibfnamefont{D.}~\bibnamefont{J.Wineland}},
  \bibinfo{journal}{Fortschr. Phys.} \textbf{\bibinfo{volume}{54}},
  \bibinfo{pages}{666} (\bibinfo{year}{2006}).

\bibitem[{\citenamefont{Pearson et~al.}(2006)\citenamefont{Pearson, Leibrandt,
  Bakr, Mallard, Brown, and Chuang}}]{Pearson}
\bibinfo{author}{\bibfnamefont{C.~E.} \bibnamefont{Pearson}},
  \bibinfo{author}{\bibfnamefont{D.~R.} \bibnamefont{Leibrandt}},
  \bibinfo{author}{\bibfnamefont{W.~S.} \bibnamefont{Bakr}},
  \bibinfo{author}{\bibfnamefont{W.~J.} \bibnamefont{Mallard}},
  \bibinfo{author}{\bibfnamefont{K.~R.} \bibnamefont{Brown}}, \bibnamefont{and}
  \bibinfo{author}{\bibfnamefont{I.~L.} \bibnamefont{Chuang}},
  \bibinfo{journal}{Phys. Rev. A} \textbf{\bibinfo{volume}{73}},
  \bibinfo{pages}{032307} (\bibinfo{year}{2006}).

\bibitem[{\citenamefont{Hucul et~al.}(2008)\citenamefont{Hucul, Yeo, Hensinger,
  Rabchuk, Olmschenk, and Monroe}}]{Hucul}
\bibinfo{author}{\bibfnamefont{D.}~\bibnamefont{Hucul}},
  \bibinfo{author}{\bibfnamefont{M.}~\bibnamefont{Yeo}},
  \bibinfo{author}{\bibfnamefont{W.~K.}~\bibnamefont{Hensinger}},
  \bibinfo{author}{\bibfnamefont{J.}~\bibnamefont{Rabchuk}},
  \bibinfo{author}{\bibfnamefont{S.}~\bibnamefont{Olmschenk}},
  \bibnamefont{and} \bibinfo{author}{\bibfnamefont{C.}~\bibnamefont{Monroe}},
  \bibinfo{journal}{Quantum Inf. Comput.} \textbf{\bibinfo{volume}{8}},
  \bibinfo{pages}{0501} (\bibinfo{year}{2008}).

\bibitem[{\citenamefont{Huber et~al.}(2008{\natexlab{a}})\citenamefont{Huber,
  Deuschle, Schnitzler, Reichle, Singer, and Schmidt-Kaler}}]{Huber}
\bibinfo{author}{\bibfnamefont{G.}~\bibnamefont{Huber}},
  \bibinfo{author}{\bibfnamefont{T.}~\bibnamefont{Deuschle}},
  \bibinfo{author}{\bibfnamefont{W.}~\bibnamefont{Schnitzler}},
  \bibinfo{author}{\bibfnamefont{R.}~\bibnamefont{Reichle}},
  \bibinfo{author}{\bibfnamefont{K.}~\bibnamefont{Singer}}, \bibnamefont{and}
  \bibinfo{author}{\bibfnamefont{F.}~\bibnamefont{Schmidt-Kaler}},
  \bibinfo{journal}{New J. Phys.} \textbf{\bibinfo{volume}{10}},
  \bibinfo{pages}{013004} (\bibinfo{year}{2008}{\natexlab{a}}).

\bibitem[{\citenamefont{Blakestad et~al.}(2009)\citenamefont{Blakestad,
  Ospelkaus, VanDevender, Amini, Britton, Leibfried, and Wineland}}]{Blakestad}
\bibinfo{author}{\bibfnamefont{R.~B.} \bibnamefont{Blakestad}},
  \bibinfo{author}{\bibfnamefont{C.}~\bibnamefont{Ospelkaus}},
  \bibinfo{author}{\bibfnamefont{A.~P.} \bibnamefont{VanDevender}},
  \bibinfo{author}{\bibfnamefont{J.~M.} \bibnamefont{Amini}},
  \bibinfo{author}{\bibfnamefont{J.}~\bibnamefont{Britton}},
  \bibinfo{author}{\bibfnamefont{D.}~\bibnamefont{Leibfried}},
  \bibnamefont{and} \bibinfo{author}{\bibfnamefont{D.~J.}
  \bibnamefont{Wineland}}, \bibinfo{journal}{Phys. Rev. Lett.}
  \textbf{\bibinfo{volume}{102}}, \bibinfo{pages}{153002}
  (\bibinfo{year}{2009}).

\bibitem[{\citenamefont{Amini et~al.}(2010)\citenamefont{Amini, Uys, Wesenberg,
  Seidelin, Britton, Bollinger, Leibfried, Ospelkaus, VanDevender, and
  Wineland}}]{Aminiz}
\bibinfo{author}{\bibfnamefont{J.~M.} \bibnamefont{Amini}},
  \bibinfo{author}{\bibfnamefont{H.}~\bibnamefont{Uys}},
  \bibinfo{author}{\bibfnamefont{J.~H.} \bibnamefont{Wesenberg}},
  \bibinfo{author}{\bibfnamefont{S.}~\bibnamefont{Seidelin}},
  \bibinfo{author}{\bibfnamefont{J.}~\bibnamefont{Britton}},
  \bibinfo{author}{\bibfnamefont{J.~J.} \bibnamefont{Bollinger}},
  \bibinfo{author}{\bibfnamefont{D.}~\bibnamefont{Leibfried}},
  \bibinfo{author}{\bibfnamefont{C.}~\bibnamefont{Ospelkaus}},
  \bibinfo{author}{\bibfnamefont{A.~P.} \bibnamefont{VanDevender}},
  \bibnamefont{and} \bibinfo{author}{\bibfnamefont{D.~J.}
  \bibnamefont{Wineland}}, \bibinfo{journal}{New J. Phys.}
  \textbf{\bibinfo{volume}{12}}, \bibinfo{pages}{033031}
  (\bibinfo{year}{2010}).

\bibitem[{\citenamefont{Stick et~al.}(2006)\citenamefont{Stick, Hensinger,
  Olmschenk, Madsen, Schwab, and Monroe}}]{stick}
\bibinfo{author}{\bibfnamefont{D.}~\bibnamefont{Stick}},
  \bibinfo{author}{\bibfnamefont{W.~K.} \bibnamefont{Hensinger}},
  \bibinfo{author}{\bibfnamefont{S.}~\bibnamefont{Olmschenk}},
  \bibinfo{author}{\bibfnamefont{M.~J.} \bibnamefont{Madsen}},
  \bibinfo{author}{\bibfnamefont{K.}~\bibnamefont{Schwab}}, \bibnamefont{and}
  \bibinfo{author}{\bibfnamefont{C.}~\bibnamefont{Monroe}},
  \bibinfo{journal}{Nature Phys.} \textbf{\bibinfo{volume}{2}},
  \bibinfo{pages}{36} (\bibinfo{year}{2006}).

\bibitem[{\citenamefont{Chiaverini et~al.}(2006)\citenamefont{Chiaverini,
  Blakestad, Britton, Jost, Langer, Leibfried, Ozeri, and
  Wineland}}]{chiaverini2}
\bibinfo{author}{\bibfnamefont{J.}~\bibnamefont{Chiaverini}},
  \bibinfo{author}{\bibfnamefont{R.} \bibnamefont{Blakestad}},
  \bibinfo{author}{\bibfnamefont{J.}~\bibnamefont{Britton}},
  \bibinfo{author}{\bibfnamefont{J.~D.} \bibnamefont{Jost}},
  \bibinfo{author}{\bibfnamefont{C.}~\bibnamefont{Langer}},
  \bibinfo{author}{\bibfnamefont{D.}~\bibnamefont{Leibfried}},
  \bibinfo{author}{\bibfnamefont{R.}~\bibnamefont{Ozeri}}, \bibnamefont{and}
  \bibinfo{author}{\bibfnamefont{D.~J.} \bibnamefont{Wineland}},
  \bibinfo{journal}{Quantum Inf. Comput.} \textbf{\bibinfo{volume}{5}},
  \bibinfo{pages}{419} (\bibinfo{year}{2006}).

\bibitem[{\citenamefont{Brownnutt et~al.}(2006)\citenamefont{Brownnutt,
  Wilpers, Gill, Thompson, and Sinclair}}]{Brownnutt}
\bibinfo{author}{\bibfnamefont{M.}~\bibnamefont{Brownnutt}},
  \bibinfo{author}{\bibfnamefont{G.}~\bibnamefont{Wilpers}},
  \bibinfo{author}{\bibfnamefont{P.}~\bibnamefont{Gill}},
  \bibinfo{author}{\bibfnamefont{R.~C.} \bibnamefont{Thompson}},
  \bibnamefont{and} \bibinfo{author}{\bibfnamefont{A.~G.}
  \bibnamefont{Sinclair}}, \bibinfo{journal}{New J. Phys.}
  \textbf{\bibinfo{volume}{8}}, \bibinfo{pages}{232} (\bibinfo{year}{2006}).

\bibitem[{\citenamefont{Brown et~al.}(2007)\citenamefont{Brown, Clark,
  Labaziewicz, Richerme, Leibrandt, and Chuang}}]{Brown}
\bibinfo{author}{\bibfnamefont{K.~R.} \bibnamefont{Brown}},
  \bibinfo{author}{\bibfnamefont{R.~J.} \bibnamefont{Clark}},
  \bibinfo{author}{\bibfnamefont{J.}~\bibnamefont{Labaziewicz}},
  \bibinfo{author}{\bibfnamefont{P.}~\bibnamefont{Richerme}},
  \bibinfo{author}{\bibfnamefont{D.~R.} \bibnamefont{Leibrandt}},
  \bibnamefont{and} \bibinfo{author}{\bibfnamefont{I.~L.}
  \bibnamefont{Chuang}}, \bibinfo{journal}{Phys. Rev. A}
  \textbf{\bibinfo{volume}{75}}, \bibinfo{pages}{015401}
  (\bibinfo{year}{2007}).

\bibitem[{\citenamefont{Britton et~al.}(2006)\citenamefont{Britton, Leibfried,
  Beall, Blakestad, Bollinger, Chiaverini, Epstein, Jost, Kielpinski, Langer
  et~al.}}]{Britton}
  \bibinfo{author}{\bibfnamefont{J.}~\bibnamefont{Britton}},
  \bibinfo{author}{\bibfnamefont{D.}~\bibnamefont{Leibfried}},
  \bibinfo{author}{\bibfnamefont{J.}~\bibnamefont{Beall}},
  \bibinfo{author}{\bibfnamefont{R.~B.} \bibnamefont{Blakestad}},
  \bibinfo{author}{\bibfnamefont{J.~J.} \bibnamefont{Bollinger}},
  \bibinfo{author}{\bibfnamefont{J.}~\bibnamefont{Chiaverini}},
  \bibinfo{author}{\bibfnamefont{R.~J.} \bibnamefont{Epstein}},
  \bibinfo{author}{\bibfnamefont{J.~D.} \bibnamefont{Jost}},
  \bibinfo{author}{\bibfnamefont{D.}~\bibnamefont{Kielpinski}},
  \bibinfo{author}{\bibfnamefont{C.}~\bibnamefont{Langer}},
  \bibinfo{author}{\bibfnamefont{R.}~\bibnamefont{Ozeri}},
   \bibinfo{author}{\bibfnamefont{R.}~\bibnamefont{Reichle}},
   \bibinfo{author}{\bibfnamefont{S.}~\bibnamefont{Seidelin}},
   \bibinfo{author}{\bibfnamefont{N.}~\bibnamefont{Shiga}},
   \bibinfo{author}{\bibfnamefont{J.~H.} \bibnamefont{Wesenberg}},
        \bibnamefont{and}
  \bibinfo{author}{\bibfnamefont{D.~J.} \bibnamefont{Wineland}}, \bibinfo{journal}{arXiv:quant-ph/0605170}
  (\bibinfo{year}{2006}).

\bibitem[{\citenamefont{Huber et~al.}(2008{\natexlab{b}})\citenamefont{Huber,
  Deuschle, Schnitzler, Reichle, Singer, and Schmidt-Kaler}}]{steane}
\bibinfo{author}{\bibfnamefont{G.}~\bibnamefont{Huber}},
  \bibinfo{author}{\bibfnamefont{T.}~\bibnamefont{Deuschle}},
  \bibinfo{author}{\bibfnamefont{W.}~\bibnamefont{Schnitzler}},
  \bibinfo{author}{\bibfnamefont{R.}~\bibnamefont{Reichle}},
  \bibinfo{author}{\bibfnamefont{K.}~\bibnamefont{Singer}}, \bibnamefont{and}
  \bibinfo{author}{\bibfnamefont{F.}~\bibnamefont{Schmidt-Kaler}},
  \bibinfo{journal}{New J. Phys.} \textbf{\bibinfo{volume}{10}},
  \bibinfo{pages}{013004} (\bibinfo{year}{2008}{\natexlab{b}}).

\bibitem[{\citenamefont{Seidelin et~al.}(2006)\citenamefont{Seidelin,
  Chiaverini, Reichle, Bollinger, Leibfried, Britton, Wesenberg, Blakestad,
  Epstein, Hume, Itano, Jost, Langer, Ozeri, Shiga, and Wineland}}]{Seidelin}
\bibinfo{author}{\bibfnamefont{S.}~\bibnamefont{Seidelin}},
  \bibinfo{author}{\bibfnamefont{J.}~\bibnamefont{Chiaverini}},
  \bibinfo{author}{\bibfnamefont{R.}~\bibnamefont{Reichle}},
  \bibinfo{author}{\bibfnamefont{J.~J.} \bibnamefont{Bollinger}},
  \bibinfo{author}{\bibfnamefont{D.}~\bibnamefont{Leibfried}},
  \bibinfo{author}{\bibfnamefont{J.}~\bibnamefont{Britton}},
  \bibinfo{author}{\bibfnamefont{J.~H.} \bibnamefont{Wesenberg}},
  \bibinfo{author}{\bibfnamefont{R.~B.} \bibnamefont{Blakestad}},
  \bibinfo{author}{\bibfnamefont{R.~J.} \bibnamefont{Epstein}},
  \bibinfo{author}{\bibfnamefont{D.~B.} \bibnamefont{Hume}},
    \bibinfo{author}{\bibfnamefont{W.~W.}~\bibnamefont{Itano}},
  \bibinfo{author}{\bibfnamefont{J.~D.} \bibnamefont{Jost}},
  \bibinfo{author}{\bibfnamefont{C.}~\bibnamefont{Langer}},
  \bibinfo{author}{\bibfnamefont{R.}~\bibnamefont{Ozeri}},
    \bibinfo{author}{\bibfnamefont{N.}~\bibnamefont{Shiga}}, \bibnamefont{and}
  \bibinfo{author}{\bibfnamefont{D.~J.} \bibnamefont{Wineland}},
  \bibinfo{journal}{Phys. Rev. Lett.}
  \textbf{\bibinfo{volume}{96}}, \bibinfo{pages}{253003}
  (\bibinfo{year}{2006}).

\bibitem[{\citenamefont{Labaziewicz
  et~al.}(2008{\natexlab{a}})\citenamefont{Labaziewicz, Ge, Antohi, Leibrandt,
  Brown, and Chuang}}]{Labaziewicz}
\bibinfo{author}{\bibfnamefont{J.}~\bibnamefont{Labaziewicz}},
  \bibinfo{author}{\bibfnamefont{Y.}~\bibnamefont{Ge}},
  \bibinfo{author}{\bibfnamefont{P.}~\bibnamefont{Antohi}},
  \bibinfo{author}{\bibfnamefont{D.}~\bibnamefont{Leibrandt}},
  \bibinfo{author}{\bibfnamefont{K.~R.} \bibnamefont{Brown}}, \bibnamefont{and}
  \bibinfo{author}{\bibfnamefont{I.~L.} \bibnamefont{Chuang}},
  \bibinfo{journal}{Phys. Rev. Lett.} \textbf{\bibinfo{volume}{100}},
  \bibinfo{pages}{013001} (\bibinfo{year}{2008}{\natexlab{a}}).

\bibitem[{\citenamefont{Britton et~al.}(2009)\citenamefont{Britton, Leibfried,
  Beall, Blakestad, Wesenberg, and Wineland}}]{Britton2}
\bibinfo{author}{\bibfnamefont{J.}~\bibnamefont{Britton}},
  \bibinfo{author}{\bibfnamefont{D.}~\bibnamefont{Leibfried}},
  \bibinfo{author}{\bibfnamefont{J.~A.} \bibnamefont{Beall}},
  \bibinfo{author}{\bibfnamefont{R.~B.} \bibnamefont{Blakestad}},
  \bibinfo{author}{\bibfnamefont{J.~H.} \bibnamefont{Wesenberg}},
  \bibnamefont{and} \bibinfo{author}{\bibfnamefont{D.~J.}
  \bibnamefont{Wineland}}, \bibinfo{journal}{Appl. Phys. Lett.}
  \textbf{\bibinfo{volume}{95}}, \bibinfo{pages}{173102}
  (\bibinfo{year}{2009}).

\bibitem[{\citenamefont{Leibrandt et~al.}(2009)\citenamefont{Leibrandt,
  Labaziewicz, Clark, Chuang, Epstein, Ospelkaus, Wesenberg, Bollinger,
  Leibfried, Wineland, Stick, Stick, Monroe, Pai, Low, Frahm, and Slusher}}]{Leibrandt}
\bibinfo{author}{\bibfnamefont{D.~R.} \bibnamefont{Leibrandt}},
  \bibinfo{author}{\bibfnamefont{J.}~\bibnamefont{Labaziewicz}},
  \bibinfo{author}{\bibfnamefont{R.~J.} \bibnamefont{Clark}},
  \bibinfo{author}{\bibfnamefont{I.~L.} \bibnamefont{Chuang}},
  \bibinfo{author}{\bibfnamefont{R.~J.} \bibnamefont{Epstein}},
  \bibinfo{author}{\bibfnamefont{C.}~\bibnamefont{Ospelkaus}},
  \bibinfo{author}{\bibfnamefont{J.~H.} \bibnamefont{Wesenberg}},
  \bibinfo{author}{\bibfnamefont{J.~J.} \bibnamefont{Bollinger}},
  \bibinfo{author}{\bibfnamefont{D.}~\bibnamefont{Leibfried}},
  \bibinfo{author}{\bibfnamefont{D.~J.} \bibnamefont{Wineland}},
  \bibinfo{author}{\bibfnamefont{D.} \bibnamefont{Stick}},
  \bibinfo{author}{\bibfnamefont{J.} \bibnamefont{Stick}},
  \bibinfo{author}{\bibfnamefont{C} \bibnamefont{Monroe}},
  \bibinfo{author}{\bibfnamefont{C.~S.} \bibnamefont{Pai}},
  \bibinfo{author}{\bibfnamefont{Y.} \bibnamefont{Low}},
  \bibinfo{author}{\bibfnamefont{R.} \bibnamefont{Frahm}},
  \bibnamefont{and}
  \bibinfo{author}{\bibfnamefont{S.~E.} \bibnamefont{Slusher}},
  \bibinfo{journal}{Quantum Inf. Comput.}
  \textbf{\bibinfo{volume}{9}}, \bibinfo{pages}{0901} (\bibinfo{year}{2009}).

\bibitem[{\citenamefont{Allcock et~al.}(2010)\citenamefont{Allcock, Sherman,
  Stacey, Burrell, Curtis, Imreh, Linke, Szwer, Webster, Steane}}]{allcock}
\bibinfo{author}{\bibfnamefont{D.~T.~C.} \bibnamefont{Allcock}},
  \bibinfo{author}{\bibfnamefont{J.~A.} \bibnamefont{Sherman}},
  \bibinfo{author}{\bibfnamefont{D.~N.} \bibnamefont{Stacey}},
  \bibinfo{author}{\bibfnamefont{A.~H.} \bibnamefont{Burrell}},
  \bibinfo{author}{\bibfnamefont{M.~J.} \bibnamefont{Curtis}},
  \bibinfo{author}{\bibfnamefont{G.}~\bibnamefont{Imreh}},
  \bibinfo{author}{\bibfnamefont{N.~M.} \bibnamefont{Linke}},
  \bibinfo{author}{\bibfnamefont{D.~J.} \bibnamefont{Szwer}},
  \bibinfo{author}{\bibfnamefont{S.~C.} \bibnamefont{Webster}},
  \bibnamefont{and}
  \bibinfo{author}{\bibfnamefont{A.~M.} \bibnamefont{Steane}}, \bibinfo{journal}{New J. Phys.}
  \textbf{\bibinfo{volume}{12}}, \bibinfo{pages}{053026}
  (\bibinfo{year}{2010}).

\bibitem[{\citenamefont{Turchette et~al.}(2000)\citenamefont{Turchette,
  Kielpinski, King, Leibfried, Meekhof, Myatt, Rowe, Sackett, Wood, Itano, Monroe, and Wineland}}]{Turchette}
\bibinfo{author}{\bibfnamefont{Q.~A.} \bibnamefont{Turchette}},
  \bibinfo{author}{\bibfnamefont{D.}~\bibnamefont{Kielpinski}},
  \bibinfo{author}{\bibfnamefont{B.~E.} \bibnamefont{King}},
  \bibinfo{author}{\bibfnamefont{D.}~\bibnamefont{Leibfried}},
  \bibinfo{author}{\bibfnamefont{D.~M.} \bibnamefont{Meekhof}},
  \bibinfo{author}{\bibfnamefont{C.~J.} \bibnamefont{Myatt}},
  \bibinfo{author}{\bibfnamefont{M.~A.} \bibnamefont{Rowe}},
  \bibinfo{author}{\bibfnamefont{C.~A.} \bibnamefont{Sackett}},
  \bibinfo{author}{\bibfnamefont{C.~S.} \bibnamefont{Wood}},
  \bibinfo{author}{\bibfnamefont{W.~M.} \bibnamefont{Itano}},
  \bibinfo{author}{\bibfnamefont{C.} \bibnamefont{Monroe}},
  \bibnamefont{and}
  \bibinfo{author}{\bibfnamefont{D.J.} \bibnamefont{Wineland}}, \bibinfo{journal}{Phys. Rev. A}
  \textbf{\bibinfo{volume}{61}}, \bibinfo{pages}{063418}
  (\bibinfo{year}{2000}).

\bibitem[{\citenamefont{Deslauriers et~al.}(2006)\citenamefont{Deslauriers,
  Olmschenk, Stick, Hensinger, Sterk, and Monroe}}]{Deslauriers2}
\bibinfo{author}{\bibfnamefont{L.}~\bibnamefont{Deslauriers}},
  \bibinfo{author}{\bibfnamefont{S.}~\bibnamefont{Olmschenk}},
  \bibinfo{author}{\bibfnamefont{D.}~\bibnamefont{Stick}},
  \bibinfo{author}{\bibfnamefont{W.~K.} \bibnamefont{Hensinger}},
  \bibinfo{author}{\bibfnamefont{J.}~\bibnamefont{Sterk}}, \bibnamefont{and}
  \bibinfo{author}{\bibfnamefont{C.}~\bibnamefont{Monroe}},
  \bibinfo{journal}{Phys. Rev. Lett.} \textbf{\bibinfo{volume}{97}},
  \bibinfo{pages}{103007} (\bibinfo{year}{2006}).

\bibitem[{\citenamefont{Labaziewicz
  et~al.}(2008{\natexlab{b}})\citenamefont{Labaziewicz, Ge, Leibrandt, Wang,
  Shewmon, and Chuang}}]{Labaziewicz2}
\bibinfo{author}{\bibfnamefont{J.}~\bibnamefont{Labaziewicz}},
  \bibinfo{author}{\bibfnamefont{Y.}~\bibnamefont{Ge}},
  \bibinfo{author}{\bibfnamefont{D.~R.} \bibnamefont{Leibrandt}},
  \bibinfo{author}{\bibfnamefont{S.~X.} \bibnamefont{Wang}},
  \bibinfo{author}{\bibfnamefont{R.}~\bibnamefont{Shewmon}}, \bibnamefont{and}
  \bibinfo{author}{\bibfnamefont{I.~L.} \bibnamefont{Chuang}},
  \bibinfo{journal}{Phys. Rev. Lett.} \textbf{\bibinfo{volume}{101}},
  \bibinfo{pages}{180602} (\bibinfo{year}{2008}{\natexlab{b}}).

\bibitem[{\citenamefont{Wesenberg et~al.}(2007)\citenamefont{Wesenberg,
  Epstein, Leibfried, Blakestad, Britton, Home, Itano, Jost, Knill, Langer, Ozeri, Seidelin, and Wineland}}]{NIST1}
\bibinfo{author}{\bibfnamefont{J.~H.} \bibnamefont{Wesenberg}},
  \bibinfo{author}{\bibfnamefont{R.~J.} \bibnamefont{Epstein}},
  \bibinfo{author}{\bibfnamefont{D.}~\bibnamefont{Leibfried}},
  \bibinfo{author}{\bibfnamefont{R.~B.} \bibnamefont{Blakestad}},
  \bibinfo{author}{\bibfnamefont{J.}~\bibnamefont{Britton}},
  \bibinfo{author}{\bibfnamefont{J.~P.} \bibnamefont{Home}},
  \bibinfo{author}{\bibfnamefont{W.~M.} \bibnamefont{Itano}},
  \bibinfo{author}{\bibfnamefont{J.~D.} \bibnamefont{Jost}},
  \bibinfo{author}{\bibfnamefont{E.}~\bibnamefont{Knill}},
  \bibinfo{author}{\bibfnamefont{C.}~\bibnamefont{Langer}},
    \bibinfo{author}{\bibfnamefont{R.}~\bibnamefont{Ozeri}},
    \bibinfo{author}{\bibfnamefont{S.}~\bibnamefont{Seidelin}}, \bibnamefont{and}
  \bibinfo{author}{\bibfnamefont{D.~J.} \bibnamefont{Wineland}}, \bibinfo{journal}{Phys. Rev. A}
  \textbf{\bibinfo{volume}{76}}, \bibinfo{pages}{053416}
  (\bibinfo{year}{2007}).

\bibitem[{\citenamefont{Epstein et~al.}(2007)\citenamefont{Epstein, Seidelin,
  Leibfried, Wesenberg, Bollinger, Amini, Blakestad, Britton, Home, Itano, Jost, Knill, Langer, Ozeri, Shiga, and Wineland}}]{NIST2}
\bibinfo{author}{\bibfnamefont{R.~J.} \bibnamefont{Epstein}},
  \bibinfo{author}{\bibfnamefont{S.}~\bibnamefont{Seidelin}},
  \bibinfo{author}{\bibfnamefont{D.}~\bibnamefont{Leibfried}},
  \bibinfo{author}{\bibfnamefont{J.~H.} \bibnamefont{Wesenberg}},
  \bibinfo{author}{\bibfnamefont{J.~J.} \bibnamefont{Bollinger}},
  \bibinfo{author}{\bibfnamefont{J.~M.} \bibnamefont{Amini}},
  \bibinfo{author}{\bibfnamefont{R.~B.} \bibnamefont{Blakestad}},
  \bibinfo{author}{\bibfnamefont{J.}~\bibnamefont{Britton}},
  \bibinfo{author}{\bibfnamefont{J.~P.} \bibnamefont{Home}},
  \bibinfo{author}{\bibfnamefont{W.~M.} \bibnamefont{Itano}},
    \bibinfo{author}{\bibfnamefont{J.~D.} \bibnamefont{Jost}},
  \bibinfo{author}{\bibfnamefont{E.}~\bibnamefont{Knill}},
  \bibinfo{author}{\bibfnamefont{C.}~\bibnamefont{Langer}},
    \bibinfo{author}{\bibfnamefont{R.}~\bibnamefont{Ozeri}},
    \bibinfo{author}{\bibfnamefont{N.}~\bibnamefont{Shiga}}, \bibnamefont{and}
  \bibinfo{author}{\bibfnamefont{D.~J.} \bibnamefont{Wineland}}, \bibinfo{journal}{Phys. Rev. A}
  \textbf{\bibinfo{volume}{76}}, \bibinfo{pages}{033411}
  (\bibinfo{year}{2007}).

\bibitem[{\citenamefont{Dietrich et~al.}(2009)\citenamefont{Dietrich, Avril,
  Bowler, Kurz, Salacka, Shu, and Blinov}}]{Dietrich}
\bibinfo{author}{\bibfnamefont{M.~R.} \bibnamefont{Dietrich}},
  \bibinfo{author}{\bibfnamefont{A.}~\bibnamefont{Avril}},
  \bibinfo{author}{\bibfnamefont{R.}~\bibnamefont{Bowler}},
  \bibinfo{author}{\bibfnamefont{N.}~\bibnamefont{Kurz}},
  \bibinfo{author}{\bibfnamefont{J.~S.} \bibnamefont{Salacka}},
  \bibinfo{author}{\bibfnamefont{G.}~\bibnamefont{Shu}}, \bibnamefont{and}
  \bibinfo{author}{\bibfnamefont{B.~B.} \bibnamefont{Blinov}},
  \bibinfo{journal}{arXiv:0905.2701}  (\bibinfo{year}{2009}).

\bibitem[{\citenamefont{Wineland
  et~al.}(1998{\natexlab{b}})\citenamefont{Wineland, Monroe, Itano, King,
  Leibfried, Meekhof, Myatt, and Wood}}]{wineland2}
\bibinfo{author}{\bibfnamefont{D.}~\bibnamefont{Wineland}},
  \bibinfo{author}{\bibfnamefont{C.}~\bibnamefont{Monroe}},
  \bibinfo{author}{\bibfnamefont{W.}~\bibnamefont{Itano}},
  \bibinfo{author}{\bibfnamefont{B.}~\bibnamefont{King}},
  \bibinfo{author}{\bibfnamefont{D.}~\bibnamefont{Leibfried}},
  \bibinfo{author}{\bibfnamefont{D.}~\bibnamefont{Meekhof}},
  \bibinfo{author}{\bibfnamefont{C.}~\bibnamefont{Myatt}}, \bibnamefont{and}
  \bibinfo{author}{\bibfnamefont{C.}~\bibnamefont{Wood}},
  \bibinfo{journal}{Fortschr. Phys.} \textbf{\bibinfo{volume}{46}},
  \bibinfo{pages}{363} (\bibinfo{year}{1998}{\natexlab{b}}).

\bibitem[{\citenamefont{N\"agerl et~al.}(1998)\citenamefont{N\"agerl, Bechter,
  Eschner, Schmidt-Kaler, and Blatt}}]{nagerl}
\bibinfo{author}{\bibfnamefont{H.}~\bibnamefont{N\"agerl}},
  \bibinfo{author}{\bibfnamefont{W.}~\bibnamefont{Bechter}},
  \bibinfo{author}{\bibfnamefont{J.}~\bibnamefont{Eschner}},
  \bibinfo{author}{\bibfnamefont{F.}~\bibnamefont{Schmidt-Kaler}},
  \bibnamefont{and} \bibinfo{author}{\bibfnamefont{R.}~\bibnamefont{Blatt}},
  \bibinfo{journal}{Appl. Phys. B} \textbf{\bibinfo{volume}{66}},
  \bibinfo{pages}{603} (\bibinfo{year}{1998}).

\bibitem[{\citenamefont{Lucas et~al.}(1998)\citenamefont{Lucas, Donald, Home,
  McDonnell, Ramos, Stacey, Stacey, Steane, and Webster}}]{Lucas}
\bibinfo{author}{\bibfnamefont{D.~M.} \bibnamefont{Lucas}},
  \bibinfo{author}{\bibfnamefont{C.~J.~S.} \bibnamefont{Donald}},
  \bibinfo{author}{\bibfnamefont{J.~P.} \bibnamefont{Home}},
  \bibinfo{author}{\bibfnamefont{M.~J.} \bibnamefont{McDonnell}},
  \bibinfo{author}{\bibfnamefont{A.}~\bibnamefont{Ramos}},
  \bibinfo{author}{\bibfnamefont{D.~N.} \bibnamefont{Stacey}},
  \bibinfo{author}{\bibfnamefont{J.-P.} \bibnamefont{Stacey}},
  \bibinfo{author}{\bibfnamefont{A.~M.} \bibnamefont{Steane}},
  \bibnamefont{and} \bibinfo{author}{\bibfnamefont{S.~C.}
  \bibnamefont{Webster}}, \bibinfo{journal}{Phil. Trans. R. Soc. A}
  \textbf{\bibinfo{volume}{66}}, \bibinfo{pages}{603} (\bibinfo{year}{1998}).

\bibitem[{\citenamefont{Hughes et~al.}(1998)\citenamefont{Hughes, James, Gomez,
  Gulley, Holzscheiter, Kwiat, Lamoreaux, Peterson, Sandberg, Schauer,
  Simmons, Thorburn, Tupa, Wang, and White}}]{hughes}
\bibinfo{author}{\bibfnamefont{R.~J.} \bibnamefont{Hughes}},
  \bibinfo{author}{\bibfnamefont{D.~F.~V.} \bibnamefont{James}},
  \bibinfo{author}{\bibfnamefont{J.~J.} \bibnamefont{Gomez}},
  \bibinfo{author}{\bibfnamefont{M.~S.} \bibnamefont{Gulley}},
  \bibinfo{author}{\bibfnamefont{M.~H.} \bibnamefont{Holzscheiter}},
  \bibinfo{author}{\bibfnamefont{P.~G.} \bibnamefont{Kwiat}},
  \bibinfo{author}{\bibfnamefont{S.~K.} \bibnamefont{Lamoreaux}},
  \bibinfo{author}{\bibfnamefont{C.~G.} \bibnamefont{Peterson}},
  \bibinfo{author}{\bibfnamefont{V.~D.} \bibnamefont{Sandberg}},
  \bibinfo{author}{\bibfnamefont{M.~M.} \bibnamefont{Schauer}},
  \bibinfo{author}{\bibfnamefont{C.~M.} \bibnamefont{Simmons}},
  \bibinfo{author}{\bibfnamefont{C.~E.} \bibnamefont{Thorburn}},
  \bibinfo{author}{\bibfnamefont{D.} \bibnamefont{Tupa}},
  \bibinfo{author}{\bibfnamefont{P.~Z.} \bibnamefont{Wang}},
  \bibnamefont{and}
  \bibinfo{author}{\bibfnamefont{A.G.} \bibnamefont{White}}, \bibinfo{journal}{Fortschr. Phys.}
  \textbf{\bibinfo{volume}{46}}, \bibinfo{pages}{329} (\bibinfo{year}{1998}).

\bibitem[{\citenamefont{Koo et~al.}(2004)\citenamefont{Koo, Sudbery, Segal, and
  Thompson}}]{thompson}
\bibinfo{author}{\bibfnamefont{K.}~\bibnamefont{Koo}},
  \bibinfo{author}{\bibfnamefont{J.}~\bibnamefont{Sudbery}},
  \bibinfo{author}{\bibfnamefont{D.~M.} \bibnamefont{Segal}}, \bibnamefont{and}
  \bibinfo{author}{\bibfnamefont{R.~C.} \bibnamefont{Thompson}},
  \bibinfo{journal}{Phys. Rev. A} \textbf{\bibinfo{volume}{69}},
  \bibinfo{pages}{043402} (\bibinfo{year}{2004}).

\bibitem[{\citenamefont{Schulz et~al.}(2008)\citenamefont{Schulz, Poschinger,
  Ziesel, and Schmidt-Kaler}}]{schulz}
\bibinfo{author}{\bibfnamefont{S.~A.} \bibnamefont{Schulz}},
  \bibinfo{author}{\bibfnamefont{U.}~\bibnamefont{Poschinger}},
  \bibinfo{author}{\bibfnamefont{F.}~\bibnamefont{Ziesel}}, \bibnamefont{and}
  \bibinfo{author}{\bibfnamefont{F.}~\bibnamefont{Schmidt-Kaler}},
  \bibinfo{journal}{New J. Phys.} \textbf{\bibinfo{volume}{73}},
  \bibinfo{pages}{045007} (\bibinfo{year}{2008}).

\bibitem[{\citenamefont{Blinov et~al.}(2002)\citenamefont{Blinov, Deslauriers,
  Lee, Madsen, Miller, and Monroe}}]{blinov}
\bibinfo{author}{\bibfnamefont{B.~B.} \bibnamefont{Blinov}},
  \bibinfo{author}{\bibfnamefont{L.}~\bibnamefont{Deslauriers}},
  \bibinfo{author}{\bibfnamefont{P.}~\bibnamefont{Lee}},
  \bibinfo{author}{\bibfnamefont{M.~J.} \bibnamefont{Madsen}},
  \bibinfo{author}{\bibfnamefont{R.}~\bibnamefont{Miller}}, \bibnamefont{and}
  \bibinfo{author}{\bibfnamefont{C.}~\bibnamefont{Monroe}},
  \bibinfo{journal}{Phys. Rev. A} \textbf{\bibinfo{volume}{65}},
  \bibinfo{pages}{040304(R)} (\bibinfo{year}{2002}).

\bibitem[{\citenamefont{Barrett et~al.}(2003)\citenamefont{Barrett, DeMarco,
  Schaetz, Meyer, Leibfried, Britton, Chiaverini, Itano, Jelenkovi\'c, Jost, Langer, Rosenbrand, and Wineland}}]{barrett}
\bibinfo{author}{\bibfnamefont{M.~D.} \bibnamefont{Barrett}},
  \bibinfo{author}{\bibfnamefont{B.}~\bibnamefont{DeMarco}},
  \bibinfo{author}{\bibfnamefont{T.}~\bibnamefont{Schaetz}},
  \bibinfo{author}{\bibfnamefont{V.}~\bibnamefont{Meyer}},
  \bibinfo{author}{\bibfnamefont{D.}~\bibnamefont{Leibfried}},
  \bibinfo{author}{\bibfnamefont{J.}~\bibnamefont{Britton}},
  \bibinfo{author}{\bibfnamefont{J.}~\bibnamefont{Chiaverini}},
  \bibinfo{author}{\bibfnamefont{W.~M.} \bibnamefont{Itano}},
  \bibinfo{author}{\bibfnamefont{B.}~\bibnamefont{Jelenkovi\'c}},
  \bibinfo{author}{\bibfnamefont{J.~D.} \bibnamefont{Jost}},
  \bibinfo{author}{\bibfnamefont{C.} \bibnamefont{Langer}},
  \bibinfo{author}{\bibfnamefont{T.}~\bibnamefont{Rosenbrand}}, \bibnamefont{and}
  \bibinfo{author}{\bibfnamefont{D.~J.}~\bibnamefont{Wineland}},
  \bibinfo{journal}{Phys. Rev. A}
  \textbf{\bibinfo{volume}{68}}, \bibinfo{pages}{042302}
  (\bibinfo{year}{2003}).

\bibitem[{\citenamefont{Letchumanan et~al.}(2007)\citenamefont{Letchumanan,
  Wilpers, Brownnutt, Gill, and Sinclair}}]{Letchumanan}
\bibinfo{author}{\bibfnamefont{V.}~\bibnamefont{Letchumanan}},
  \bibinfo{author}{\bibfnamefont{G.}~\bibnamefont{Wilpers}},
  \bibinfo{author}{\bibfnamefont{M.}~\bibnamefont{Brownnutt}},
  \bibinfo{author}{\bibfnamefont{P.}~\bibnamefont{Gill}}, \bibnamefont{and}
  \bibinfo{author}{\bibfnamefont{A.~G.} \bibnamefont{Sinclair}},
  \bibinfo{journal}{Phys. Rev. A} \textbf{\bibinfo{volume}{75}},
  \bibinfo{pages}{063425} (\bibinfo{year}{2007}).

\bibitem[{\citenamefont{Bell et~al.}(1991)\citenamefont{Bell, Gill, Klein, Levick, Tamm, and Schnier}}]{bell}
\bibinfo{author}{\bibfnamefont{A.~S.} \bibnamefont{Bell}},
  \bibinfo{author}{\bibfnamefont{P.}~\bibnamefont{Gill}},
  \bibinfo{author}{\bibfnamefont{H.~A.} \bibnamefont{Klein}},
  \bibinfo{author}{\bibfnamefont{A.~P.} \bibnamefont{Levick}},
    \bibinfo{author}{\bibfnamefont{C.} \bibnamefont{Tamm}}, \bibnamefont{and}
      \bibinfo{author}{\bibfnamefont{D.} \bibnamefont{Schnier}},
  \bibinfo{journal}{Phys. Rev. A} \textbf{\bibinfo{volume}{44}},
  \bibinfo{pages}{20} (\bibinfo{year}{1991}).

\bibitem[{\citenamefont{Balzer et~al.}(2006)\citenamefont{Balzer, Braun,
  Hannemann, Paape, Ettler, Neuhauser, and Wunderlich}}]{balzer}
\bibinfo{author}{\bibfnamefont{C.}~\bibnamefont{Balzer}},
  \bibinfo{author}{\bibfnamefont{A.}~\bibnamefont{Braun}},
  \bibinfo{author}{\bibfnamefont{T.}~\bibnamefont{Hannemann}},
  \bibinfo{author}{\bibfnamefont{C.}~\bibnamefont{Paape}},
  \bibinfo{author}{\bibfnamefont{M.}~\bibnamefont{Ettler}},
  \bibinfo{author}{\bibfnamefont{W.}~\bibnamefont{Neuhauser}},
  \bibnamefont{and}
  \bibinfo{author}{\bibfnamefont{C.}~\bibnamefont{Wunderlich}},
  \bibinfo{journal}{Phys. Rev. A} \textbf{\bibinfo{volume}{73}},
  \bibinfo{pages}{041407(R)} (\bibinfo{year}{2006}).

\bibitem[{\citenamefont{Kielpinski et~al.}(2006)\citenamefont{Kielpinski,
  Cetina, Cox, and K\"artner}}]{kielpinski}
\bibinfo{author}{\bibfnamefont{D.}~\bibnamefont{Kielpinski}},
  \bibinfo{author}{\bibfnamefont{M.}~\bibnamefont{Cetina}},
  \bibinfo{author}{\bibfnamefont{J.~A.} \bibnamefont{Cox}}, \bibnamefont{and}
  \bibinfo{author}{\bibfnamefont{F.~X.} \bibnamefont{K\"artner}},
  \bibinfo{journal}{Opt. Lett.} \textbf{\bibinfo{volume}{31}},
  \bibinfo{pages}{757} (\bibinfo{year}{2006}).

\bibitem[{\citenamefont{Olmschenk et~al.}(2007)\citenamefont{Olmschenk, Younge,
  Moehring, Matsukevich, Maunz, and Monroe}}]{olmschenk}
\bibinfo{author}{\bibfnamefont{S.}~\bibnamefont{Olmschenk}},
  \bibinfo{author}{\bibfnamefont{K.~C.} \bibnamefont{Younge}},
  \bibinfo{author}{\bibfnamefont{D.~L.} \bibnamefont{Moehring}},
  \bibinfo{author}{\bibfnamefont{D.~N.} \bibnamefont{Matsukevich}},
  \bibinfo{author}{\bibfnamefont{P.}~\bibnamefont{Maunz}}, \bibnamefont{and}
  \bibinfo{author}{\bibfnamefont{C.}~\bibnamefont{Monroe}},
  \bibinfo{journal}{Phys. Rev. A} \textbf{\bibinfo{volume}{76}},
  \bibinfo{pages}{052314} (\bibinfo{year}{2007}).

\bibitem[{\citenamefont{Roberts et~al.}(1999)\citenamefont{Roberts, Taylor,
  Gateva-Kostova, Clarke, Rowley, and Gill}}]{roberts}
\bibinfo{author}{\bibfnamefont{M.}~\bibnamefont{Roberts}},
  \bibinfo{author}{\bibfnamefont{P.}~\bibnamefont{Taylor}},
  \bibinfo{author}{\bibfnamefont{S.~V.} \bibnamefont{Gateva-Kostova}},
  \bibinfo{author}{\bibfnamefont{R.~B.~M.} \bibnamefont{Clarke}},
  \bibinfo{author}{\bibfnamefont{W.~R.~C.} \bibnamefont{Rowley}},
  \bibnamefont{and} \bibinfo{author}{\bibfnamefont{P.}~\bibnamefont{Gill}},
  \bibinfo{journal}{Phys. Rev. A} \textbf{\bibinfo{volume}{60}},
  \bibinfo{pages}{2867} (\bibinfo{year}{1999}).

\bibitem[{\citenamefont{Gill et~al.}(2003)\citenamefont{Gill, Barwood, Klein,
  Huang, Webster, Blythe, Hosaka, Lea, and Margolis}}]{gill2}
\bibinfo{author}{\bibfnamefont{P.}~\bibnamefont{Gill}},
  \bibinfo{author}{\bibfnamefont{G.~P.} \bibnamefont{Barwood}},
  \bibinfo{author}{\bibfnamefont{H.~A.} \bibnamefont{Klein}},
  \bibinfo{author}{\bibfnamefont{G.}~\bibnamefont{Huang}},
  \bibinfo{author}{\bibfnamefont{S.~A.} \bibnamefont{Webster}},
  \bibinfo{author}{\bibfnamefont{P.~J.} \bibnamefont{Blythe}},
  \bibinfo{author}{\bibfnamefont{K.}~\bibnamefont{Hosaka}},
  \bibinfo{author}{\bibfnamefont{S.~N.} \bibnamefont{Lea}}, \bibnamefont{and}
  \bibinfo{author}{\bibfnamefont{H.~S.} \bibnamefont{Margolis}},
  \bibinfo{journal}{Meas. Sci. Technol.} \textbf{\bibinfo{volume}{14}},
  \bibinfo{pages}{1174}  (\bibinfo{year}{2003}).

\bibitem[{\citenamefont{Schneider et~al.}(2005)\citenamefont{Schneider, Peik,
  and Tamm}}]{schneider}
\bibinfo{author}{\bibfnamefont{T.}~\bibnamefont{Schneider}},
  \bibinfo{author}{\bibfnamefont{E.}~\bibnamefont{Peik}}, \bibnamefont{and}
  \bibinfo{author}{\bibfnamefont{C.}~\bibnamefont{Tamm}},
  \bibinfo{journal}{Phys. Rev. Lett.} \textbf{\bibinfo{volume}{94}},
  \bibinfo{pages}{230801} (\bibinfo{year}{2005}).

\bibitem[{\citenamefont{Hosaka et~al.}(2005)\citenamefont{Hosaka, Webster,
  Blythe, Stannard, Beaton, Margolis, Lea, and Gill}}]{kazumoto}
\bibinfo{author}{\bibfnamefont{K.}~\bibnamefont{Hosaka}},
  \bibinfo{author}{\bibfnamefont{S.~A.} \bibnamefont{Webster}},
  \bibinfo{author}{\bibfnamefont{P.~J.} \bibnamefont{Blythe}},
  \bibinfo{author}{\bibfnamefont{A.}~\bibnamefont{Stannard}},
  \bibinfo{author}{\bibfnamefont{D.}~\bibnamefont{Beaton}},
  \bibinfo{author}{\bibfnamefont{H.~S.} \bibnamefont{Margolis}},
  \bibinfo{author}{\bibfnamefont{S.~N.} \bibnamefont{Lea}}, \bibnamefont{and}
  \bibinfo{author}{\bibfnamefont{P.}~\bibnamefont{Gill}},
  \bibinfo{journal}{IEEE Trans. Instr. Meas.} \textbf{\bibinfo{volume}{54}},
  \bibinfo{pages}{759} (\bibinfo{year}{2005}).

\bibitem[{\citenamefont{Streed et~al.}(2008)\citenamefont{Streed, Weinhold, and
  Kielpinski}}]{streed}
\bibinfo{author}{\bibfnamefont{E.~W.} \bibnamefont{Streed}},
  \bibinfo{author}{\bibfnamefont{T.~J.} \bibnamefont{Weinhold}},
  \bibnamefont{and}
  \bibinfo{author}{\bibfnamefont{D.}~\bibnamefont{Kielpinski}},
  \bibinfo{journal}{Appl. Phys. Lett.} \textbf{\bibinfo{volume}{93}},
  \bibinfo{pages}{071103} (\bibinfo{year}{2008}).

\bibitem[{\citenamefont{Macalpine and Schildknecht}(1959)}]{Macalpiner}
\bibinfo{author}{\bibfnamefont{W.~W.} \bibnamefont{Macalpine}}
  \bibnamefont{and} \bibinfo{author}{\bibfnamefont{R.~O.}
  \bibnamefont{Schildknecht}}, \bibinfo{journal}{Pro. IRE}
  \textbf{\bibinfo{volume}{47}}, \bibinfo{pages}{2099} (\bibinfo{year}{1959}).

\bibitem[{\citenamefont{Siverns et~al.}(2010)\citenamefont{Siverns, Weidt, and
  Hensinger}}]{siverns}
\bibinfo{author}{\bibfnamefont{J.~D.} \bibnamefont{Siverns}},
  \bibinfo{author}{\bibfnamefont{S.}~\bibnamefont{Weidt}}, \bibnamefont{and}
  \bibinfo{author}{\bibfnamefont{W.~K.} \bibnamefont{Hensinger}},
  \bibinfo{journal}{in preparation}.

\bibitem[{\citenamefont{Madsen et~al.}(2004)\citenamefont{Madsen, Hensinger,
  Stick, Rabchuk, and Monroe}}]{Madsen}
\bibinfo{author}{\bibfnamefont{M.~J.} \bibnamefont{Madsen}},
  \bibinfo{author}{\bibfnamefont{W.~K.} \bibnamefont{Hensinger}},
  \bibinfo{author}{\bibfnamefont{D.}~\bibnamefont{Stick}},
  \bibinfo{author}{\bibfnamefont{J.~A.} \bibnamefont{Rabchuk}},
  \bibnamefont{and} \bibinfo{author}{\bibfnamefont{C.}~\bibnamefont{Monroe}},
  \bibinfo{journal}{Appl. Phys. B} \textbf{\bibinfo{volume}{78}},
  \bibinfo{pages}{639} (\bibinfo{year}{2004}).

\bibitem[{\citenamefont{Olmschenk et~al.}(2009)\citenamefont{Olmschenk, Hayes,
  Matsukevich, Maunz, Moehring, Younge, and Monroe}}]{Olmschenk3}
\bibinfo{author}{\bibfnamefont{S.}~\bibnamefont{Olmschenk}},
  \bibinfo{author}{\bibfnamefont{D.}~\bibnamefont{Hayes}},
  \bibinfo{author}{\bibfnamefont{D.~N.} \bibnamefont{Matsukevich}},
  \bibinfo{author}{\bibfnamefont{P.}~\bibnamefont{Maunz}},
  \bibinfo{author}{\bibfnamefont{D.~L.} \bibnamefont{Moehring}},
  \bibinfo{author}{\bibfnamefont{K.~C.} \bibnamefont{Younge}},
  \bibnamefont{and} \bibinfo{author}{\bibfnamefont{C.}~\bibnamefont{Monroe}},
  \bibinfo{journal}{Phys. Rev. A} \textbf{\bibinfo{volume}{80}},
  \bibinfo{pages}{022502} (\bibinfo{year}{2009}).

\bibitem[{\citenamefont{Engelke and Tamm}(1996)}]{Engelke}
\bibinfo{author}{\bibfnamefont{D.}~\bibnamefont{Engelke}} \bibnamefont{and}
  \bibinfo{author}{\bibfnamefont{C.}~\bibnamefont{Tamm}},
  \bibinfo{journal}{Europhys. Lett.} \textbf{\bibinfo{volume}{33}},
  \bibinfo{pages}{347} (\bibinfo{year}{1996}).

\bibitem[{\citenamefont{Zhao et~al.}(1998)\citenamefont{Zhao, Simsarian,
  Orozco, and Sprouse}}]{comp-based}
\bibinfo{author}{\bibfnamefont{W.~Z.} \bibnamefont{Zhao}},
  \bibinfo{author}{\bibfnamefont{J.~E.} \bibnamefont{Simsarian}},
  \bibinfo{author}{\bibfnamefont{L.~A.} \bibnamefont{Orozco}},
  \bibnamefont{and} \bibinfo{author}{\bibfnamefont{G.~D.}
  \bibnamefont{Sprouse}}, \bibinfo{journal}{Rev. Sci. Instr.}
  \textbf{\bibinfo{volume}{69}}, \bibinfo{pages}{3737} (\bibinfo{year}{1998}).

\bibitem[{\citenamefont{Nizamani et~al.}(2010)\citenamefont{Nizamani,
  McLoughlin, and Hensinger}}]{Nizamani}
\bibinfo{author}{\bibfnamefont{A.~H.} \bibnamefont{Nizamani}},
  \bibinfo{author}{\bibfnamefont{J.~J.} \bibnamefont{McLoughlin}},
  \bibnamefont{and} \bibinfo{author}{\bibfnamefont{W.~K.}
  \bibnamefont{Hensinger}}, \bibinfo{journal}{Phys. Rev. A} \textbf{\bibinfo{volume}{82}}, \bibinfo{pages}{043408}
  (\bibinfo{year}{2010}).

  \bibitem[{\citenamefont{Amini, Britton, Leibfried, and Wineland}(1999)}]{Amini2}
\bibinfo{author}\bibinfo{author}{\bibfnamefont{J.~M.} \bibnamefont{Amini}},
  \bibinfo{author}{\bibfnamefont{J.}~\bibnamefont{Britton}},
  \bibinfo{author}{\bibfnamefont{D.}~\bibnamefont{Leibfried}},
  \bibnamefont{and} \bibinfo{author}{\bibfnamefont{D.~J.}
  \bibnamefont{Wineland}},
  \bibinfo{journal}{in \emph{Atom Chips}},
  \bibinfo{editor}{edited by J. Reichel and V. Vuletic}
  (\bibinfo{publisher}{WILEY-VCH},
  \bibinfo{year}{in press}).

\end{thebibliography}
\end{document}